\documentclass[journal]{IEEEtran}
\usepackage{pdfpages}

\usepackage{amsmath}
\usepackage{amssymb}
\usepackage{algorithmic}
\usepackage{algorithm}
\usepackage{enumerate}
\usepackage{graphicx}
\usepackage{color}
\usepackage{multirow,tabularx}
\usepackage{enumerate}
\usepackage{times}
\usepackage{booktabs}
\usepackage{subfloat}
\ifCLASSOPTIONcompsoc
  \usepackage[caption=false,font=footnotesize,labelfont=sf,textfont=sf]{subfig}
\else
  \usepackage[caption=false,font=footnotesize]{subfig}
\fi

\usepackage{fixltx2e}
\usepackage{dblfloatfix}
\usepackage{url}
\usepackage{bm}
\usepackage{multirow,tabularx}
\usepackage[subfigure]{tocloft}

\newcolumntype{C}{>{\centering\arraybackslash}X}

\usepackage{amsthm}

\theoremstyle{plain}

\theoremstyle{definition}

\usepackage[breaklinks=true,bookmarks=false]{hyperref}

\renewcommand{\algorithmicrequire}{\textbf{Input:}}
\renewcommand{\algorithmicensure}{\textbf{Output:}}

\ifCLASSINFOpdf
\else
\fi
\begin{document}
\title{ Deep Anti-aliasing of Whole Focal Stack Using Slice Spectrum }

\author{Yaning Li,
        Xue Wang,
        Hao Zhu,
        Guoqing Zhou, and %~\IEEEmembership{Member,~IEEE},% <-this % stops a space
        Qing Wang~\IEEEmembership{Senior Member,~IEEE}
\IEEEcompsocitemizethanks{\IEEEcompsocthanksitem Y. Li, X. Wang, G. Zhou, Q. Wang (corresponding author) are with the School of Computer Science, Northwestern Polytechnical University, Xi'an 710072, China. E-mail: qwang@nwpu.edu.cn.

H. Zhu is with the School of Electronic Science and Engineering, Nanjing University, Nanjing 210023, China. E-mail: zhuhao\_photo@nju.edu.cn.

\IEEEcompsocthanksitem The work was supported by NSFC under Grant 62031023 and Grant 61801396.}% <-this % stops an unwanted space
\thanks{Manuscript received April 25, 2021; }}


\markboth{Journal of \ transactions on computational imaging ,~Vol., No., April~2021}%
{Shell \MakeLowercase{\textit{{\it et al.}.}}: Bare Demo of IEEEtran.cls for Computer Society Journals}



\IEEEtitleabstractindextext{%
\begin{abstract}
\sloppy{}
 %the anti-aliasing issue for digital refocusing of 3D light field with horizontal- or vertical-dimension under-sampling. 
The paper aims at removing the aliasing effects of the whole focal stack generated from a sparse-sampled {4D} light field, while keeping the consistency across all the focal layers. We first explore the structural characteristics embedded in the focal stack slice and its corresponding frequency-domain representation, i.e., the Focal Stack Spectrum (FSS). 
We observe that the energy distribution of the FSS always resides within the same triangular area under different angular sampling rates, additionally the continuity of the Point Spread Function (PSF) is intrinsically maintained in the FSS. 
%We then observe that the continuity of the point spread function (PSF) is intrinsically maintained in the FSS, which enables the structural consistency along the focal direction under different angular sampling rates. 
Based on these two observations, we propose a learning-based FSS reconstruction approach for one-time aliasing removing over the whole focal stack. Moreover, a novel conjugate-symmetric loss function is proposed for the optimization. Compared to previous works, our method avoids an explicit depth estimation, and can handle challenging large-disparity scenarios. Experimental results on both synthetic and real light field datasets show the superiority of the proposed approach for different scenes and various angular sampling rates.

%Due to the unavoidable spatio-angular trade-off for light field acquisitions, digital refocused images generated from angularly undersampled light fields contain significant aliasing artifacts. Previous works perform anti-aliasing by applying depth filters, view interpolation, or multi-scale fusion. These methods either rely on accurate depth estimations, or lack the adaptivity to remove changing aliasing effects along the focal direction. In this paper, we exploit the structural characteristics embedding in the focal stack and its corresponding frequency-domain representation, the Focal Stack Spectrum (FSS). The continuity of the Point Spread Function (PSF) is intrinsically maintained in the FSS, which enables the structural consistency along the focal direction under different angular sampling rates. Based on these advantages, we propose a learning-based FSS reconstruction approach for one-time aliasing removing over the whole focal stack. What's more, a novel origin-symmetric loss function is proposed for optimization. Our method avoids an explicit depth estimation, and can handle challenging large-disparity scenarios. Experimental results on both synthetic and real light field datasets show the superiority of the proposed approach for different scenes and various angular sampling rates.
\end{abstract}

}

\maketitle

\IEEEdisplaynontitleabstractindextext

% \fi
\IEEEpeerreviewmaketitle

\IEEEraisesectionheading{}

%%%%%%%%% BODY TEXT

%\begin{figure}[!t]
%	\begin{center}
%		\centering
%		\includegraphics[width=\linewidth]{slice_focalstack.png}
%	\end{center}
%	\caption{ The differences of Anti-aliasing and aliasing effect on single focal slice, focal stack and focal stack spectrum. The first line are refocus image at  a certain depth, the second line are the focal stacks of a certain row of the image and third line is the corresponding spectrum of focal stack. The dataset are obtained with the viewpoint changing along the $u$ direction. The number of viewponts is $1\times 9 $ (a) Input data with aliasing. (b) Anti-aliasing output of our method.}
%	\label{fig:slice_focalstack}
%\end{figure}

\section{Introduction}
\sloppy{}

\begin{figure}[!t]
	\begin{center}
		
	\includegraphics[width=\linewidth]{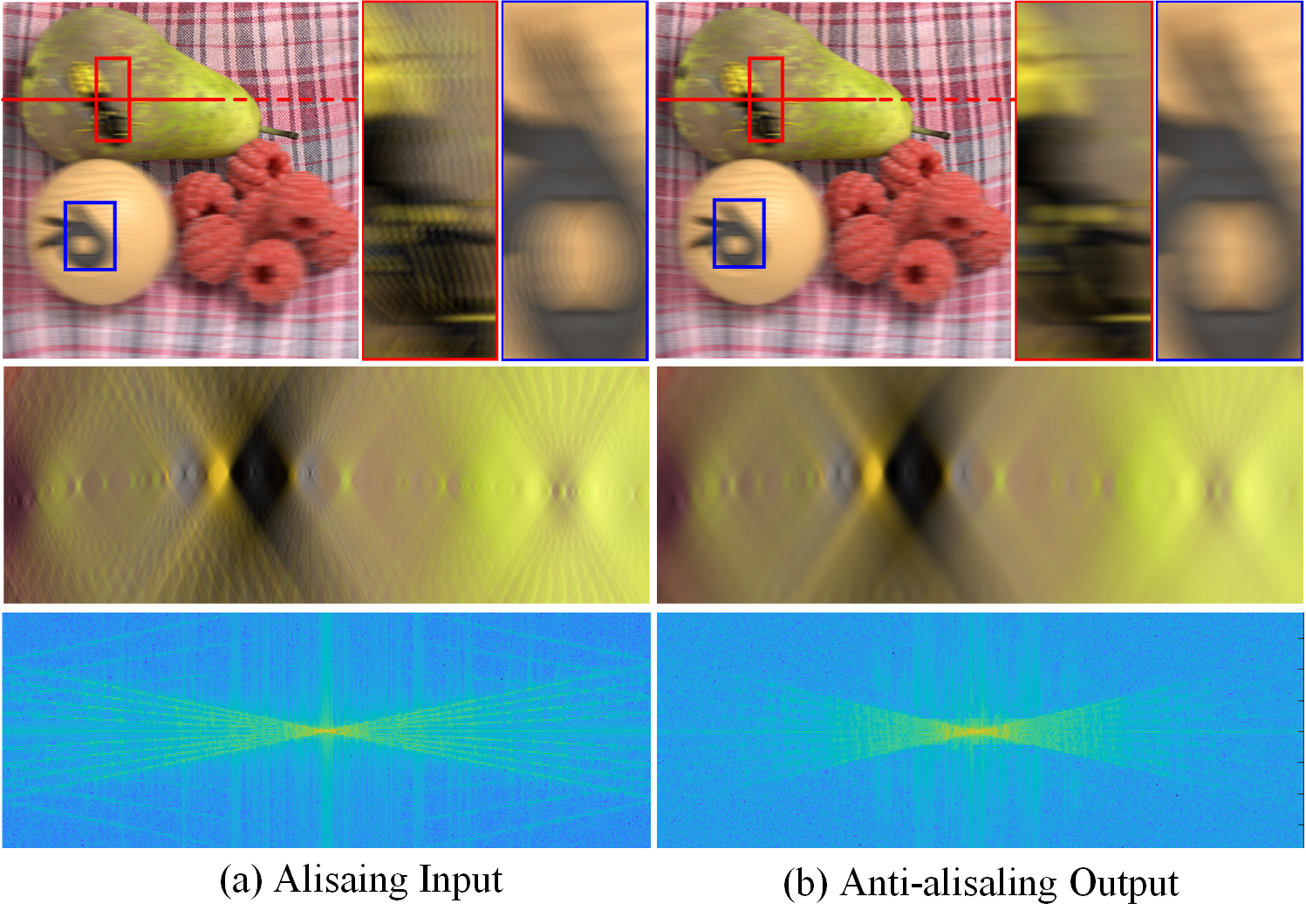}

	\end{center}

	\caption[example]
	{Aliasing effect and aliasing-removed result. % on single focal slice, focal stack and FSS. 
(a) Input with aliasing and the view count is $1\times 9$. (b) Anti-aliasing output by our method. From top to bottom, refocused image at a certain depth, focal stack slice along the red line, and corresponding FSS. \label{fig:anti_aliasing_hci_data}}
%\vspace{-1.5em}
\end{figure}

\IEEEPARstart{L}IGHT field imaging \cite{isaksen2000dynamically} enables digital refocusing at different focal planes after the time of capture. Basically this is performed by integrating a light field over the angular domain, which corresponds to the slice operation in the frequency domain \cite{ng2006digital}. However, with a sparse angular sampling, {\textit{i.e.}}, the disparity between adjacent views is more than one pixel \cite{chai2000plenoptic}, there will be significant aliasing artifacts in the out-of-focus regions in the refocused images \cite{xiao2017aliasing}, as shown in Fig.\ref{fig:anti_aliasing_hci_data}(a).

To enhance visual quality, many approaches  \cite{ xiao2017aliasing, georgiev2010reducing, levoy1996light, chang2014anti, lumsdaine2008full, dansereau2015linear,ben2020deep} have been proposed to remove the aliasing effects based on view interpolation \cite{chang2014anti,kalantari2016learning,wu2017light_epi}, depth-based filtering \cite{chai2000plenoptic}, or multi-scale fusion \cite{xiao2017aliasing}. However, since most of these methods rely on depth estimation \cite{Lin2015Depth,Broad2016Light}, inaccurate depth maps will cause severe degradation in anti-aliasing performance. Moreover, existing methods only consider an individual refocused image, which is corresponding to a specific depth layer in the whole focal stack. Without taking all layers together into consideration, they could not provide consistent enhancement over the focal stack (as shown in Fig.\ref{symmetry_property_result}). 
%\textcolor{red}{(Fig 1 consists of aliasing input and the output by our method, which dose not support the statement very well)}. 
Namely, along the axial direction in the focal stack, the PSF-continuity can not be maintained well. This will become more critical for the refocused images with large disparities or complex occlusions.

In the paper, we focus on exploring the structural characteristics embedding in the focal stack and its corresponding Fourier spectrum, named Focal Stack Spectrum (FSS). Different from the EPI (a 2D representation for a light field) where the slopes of the EPI lines vary with depths, for a given light field, the FSSs for different depths share \textbf{the same cone-shaped pattern} (as shown in the Fig.\ref{deffer_samlpy_of_efs}). In other words, the energy distribution of the FSS locates within the same triangular area. Furthermore, the PSF-continuity is intrinsically maintained in the FSS. These important characteristics of the FSS make it possible to exploit a unified anti-aliasing scheme for whole depth contents.
The main contributions of the paper are,

1) Two important characteristics of the frequency-domain representation for the light field focal stack are explored. The FSS can preserve the PSF-continuity and provide the same bounds of spectral support along the focal axis under different angular sampling rates.

2) A deep FSS-based anti-aliasing algorithm is proposed to perform one-time aliasing removal for all the refocused layers and meanwhile preserve the consistency across focal layers, only a rough disparity range estimation being needed.

3) A novel and robust conjugate-symmetric loss function is adopted in the U-Net for optimization.

\section{Related Work}
\label{Related Work}
\sloppy{}
\subsection{Light Field Refocusing}
\sloppy{}
%%The 4D light field $L(u,v,x,y)$ \cite{levoy1996light,gortler1996lumigraph} %redundantly
%records light rays in a 3D space. So far many researches have been done for analyzing the characteristics of the 4D light field sampling and digital refocusing. Ng \cite{ng2005fourier} pointed out that the spectrum of a light field concentrated on a 3D manifold and each focal image could be synthesised by applying an inverse Fourier transform to a 2D slice in the manifold. Shi {\it et al.} \cite{shi2014light} found the sparsity of a light field from which a dense light field is reconstructed by applying the sparse Fourier transform. Considering the regular 2D mesh sampling structure of the light field, Levin and Durand \cite{levin2010linear} analysed the dimensional gap between the 3D focal stack and 4D light field and proposed to inpaint the light field spectrum from the focal stack. Isaksen {\it et al.} \cite{isaksen2000dynamically} re-parameterized the light field as a 3D focal stack in the spatial domain. Dansereau {\it et al.} \cite{dansereau2015linear} extended the capability of refocus from one single depth layer to a volumetric range of depths by replacing the slice operation with a depth-dependent band-pass filter.

 The 4D light field $L(u,v,x,y)$ \cite{levoy1996light,gortler1996lumigraph}  records light rays in a 3D space, where $(u,v)$ and $(x,y)$ denote the ray's intersections with the angular and spatial planes, respectively. So far, many studies have been done for analysing the 4D light field sampling characteristics and digital refocusing. In the spatial domain, by re-parameterizing the light rays and integrating them along the angular dimensions, the scene can be refocused \cite{isaksen2000dynamically}. In the Fourier domain, Ng \cite{ng2005fourier} points out that the spectrum of a light field concentrates on a 3D manifold and each focal image could be synthesized by applying an inverse Fourier transform to a 2D slice in the manifold. After that, Dansereau {\it et al.} \cite{dansereau2015linear} extend the capability of refocus from a single depth layer to a volumetric range of depths by replacing the slice operation with a depth-dependent band-pass filter. It is worth noting that angular undersampling will cause severe aliasing artifacts and the above-mentioned methods can not remove this kind of aliasing.

%
%However, due to the limitation of sensor size, in large disparity, complex occlusion scene, there always exists a contradiction between the spatial and angular resolutions, which results in the aliasing artifacts in digital refocusing for angularly undersampled light fields \cite{ng2006digital,stewart2003new}.

\subsection{Anti-aliasing for Light Field Refocusing}
\sloppy{}
%Considering the formation of aliasing effects, it is a straightforward way to tackle anti-aliasing by angular super-resolution \cite{kalantari2016learning,wang2017light,wu2017light_epi}.
Anti-aliasing usually requires either abundant angular sampling or appropriate filters. This problem has been widely studied in both spatial and frequency domains.

%All these methods can be categorized into spatial domain and frequency domain.

\textbf{Spatial-domain methods.} Most previous approaches rely on light field prefiltering. Levoy and Hanrahan \cite{levoy1996light} employ a prefiltering to reduce the spatial artifacts. However, the prefiltering inevitably introduces over-smoothness in the focused areas. In order to mitigate the over-smoothness issue, some depth-based methods are proposed to remove the aliasing. Levin {\it et al.} \cite{2008Understanding} propose to use the mixture-of-Gaussians derivative priors to recover a nearly aliasing-free light field given the scene depths. Chang {\it et al.} \cite{chang2014anti} propose an anti-aliasing algorithm by interpolating angular sampling within each sampling interval using depth information. Lin {\it et al.} \cite{Lin2015Depth} analyse the symmetry characteristic of the focal stack slice in the spatial domain and prove it is possible to use depth-based light field rendering to reduce the aliasing. All these methods need accurate scene depths. 
Recently, with the development of light field reconstruction techniques \cite{ wu2017light_epi, yeung2018fast, mildenhall2019local, srinivasan2019pushing}, many light field angular super-resolution methods have been proposed to reduce the aliasing. Kalantari {\it {et al.}} \cite{kalantari2016learning} propose two convolutional neural networks to estimate the depth and color of each viewpoint sequentially. Wu {\it et al.} \cite{wu2017light_epi} remove the aliasing by performing angular super-resolution on the EPI. Yeung {\it et al.} \cite{yeung2018fast} propose a learning-based algorithm to reconstruct a densely-sampled LF rapidly and accurately from a sparsely-sampled LF in one forward pass. Srinivasan1 {\it et al.} \cite{srinivasan2019pushing} predict the Multiplane Image (MPI) scene representation from a narrow baseline stereo pair, which can be used for view synthesis. Mildenhall {\it et al.} \cite{mildenhall2019local} propose to render novel views by blending adjacent local light fields. These techniques require solving the scene reconstruction problem, which is a traditional challenge. Without reconstruction, Bishop {\it et al.} \cite{bishop2011light} eliminate the aliasing by fusing multiview information. By analysing the angular aliasing model in the spatial domain, Xiao {\it et al.}  \cite{xiao2017aliasing} first detect the aliasing contents and then use lower-frequency terms of decomposition to remove the angular aliasing at the refocusing stage. Dayan {\it et al.} \cite{ben2020deep} propose a convolutional neural network to remove the aliasing effects from a sparse light field. 

%
%Chang {\it et al.}  \cite{chang2014anti} proposed an anti-aliasing algorithm by interpolating angular samples within each sampling interval using depth information. Bishop {\it et al.} \cite{bishop2011light} eliminated the aliasing by fusing multiview information. Xiao {\it et al.} \cite{xiao2017aliasing} further analysed the aliasing in the spatial domain. They first detected aliasing and then used a multi-scale fusion method to remove the aliasing. Lin {\it et al.} \cite{Lin2015Depth} analysed the symmetry characteristic of focal stack slice in the spatial domain and proved it was possible to use depth-based light field rendering to reduce aliasing. Dayan \cite{ben2020deep} pro

\textbf{Frequency-domain methods.} %Unlike local characteristics utilized in spatial-domain methods, global processing are mostly taken into consideration in frequency-domain methods. 
 Isaksen {\it et al.} \cite{isaksen2000dynamically}  propose a frequency-planar light field filter. Chai {\it et al.} \cite{chai2000plenoptic} conduct a comprehensive analysis on the trade-off between sampling density and depth resolution in the frequency domain and reconstruct the EPI spectral using depth filters. Ng \cite{ng2005fourier} suggests that band-limited filtering in the frequency domain and slicing can effectively inhibit the aliasing effects. In terms of view reconstruction, Le Pendu {\it et al.} \cite{pendu2019fourier,le2020high} present a Fourier Disparity Layer (FDL) representation for light fields. Once the layers are known, they can be simply shifted and filtered to produce different viewpoints to remove the aliasing. Shi {\it et al.} \cite{shi2014light} propose to complete the angular spectrum of a light field in the continuous Fourier domain. Vagharshakyan {\it et al.} \cite{vagharshakyan2018light} iteratively compensate the high frequency spectrum of a sparse EPI representation in shearlet domain.

However, all these methods have their specific limitations. Prefiltering techniques can eliminate aliasing only while the focused areas are also over-smoothed. Depth and view reconstruction based techniques are prone to depth or reconstruction errors. When the refocused depth is far away from the original focused depth, the aliasing is aggravated. Moreover, all these methods tackle each refocused image individually so that they can not provide a PSF-continuous aliasing-removed focal stack (as shown in  Fig.\ref{symmetry_property_result}). Different from these methods, the proposed FSS representation enables the same cone-shaped distribution pattern in the frequency domain shared by different scenarios, which provides a unified depth-independent solution for generating the PSF-continuous anti-aliasing focal stack in one single pass.

%-------------------------------------------------------------------------
\section{Focal Stack Spectrum}\label{sec:FSS}
\sloppy{}
In this section, we first elucidate the way to obtain the FSS from a light field, and then analyse the characteristics of the FSS. Without loss of generality, a 2D EPI instead of a full 4D one is used here to deliver good demonstrations.

\subsection{Definition of FSS}
\label{sec:subFSS}
\sloppy{}
\begin{figure*}[!tb]
	\centering
	\includegraphics[width=\linewidth]{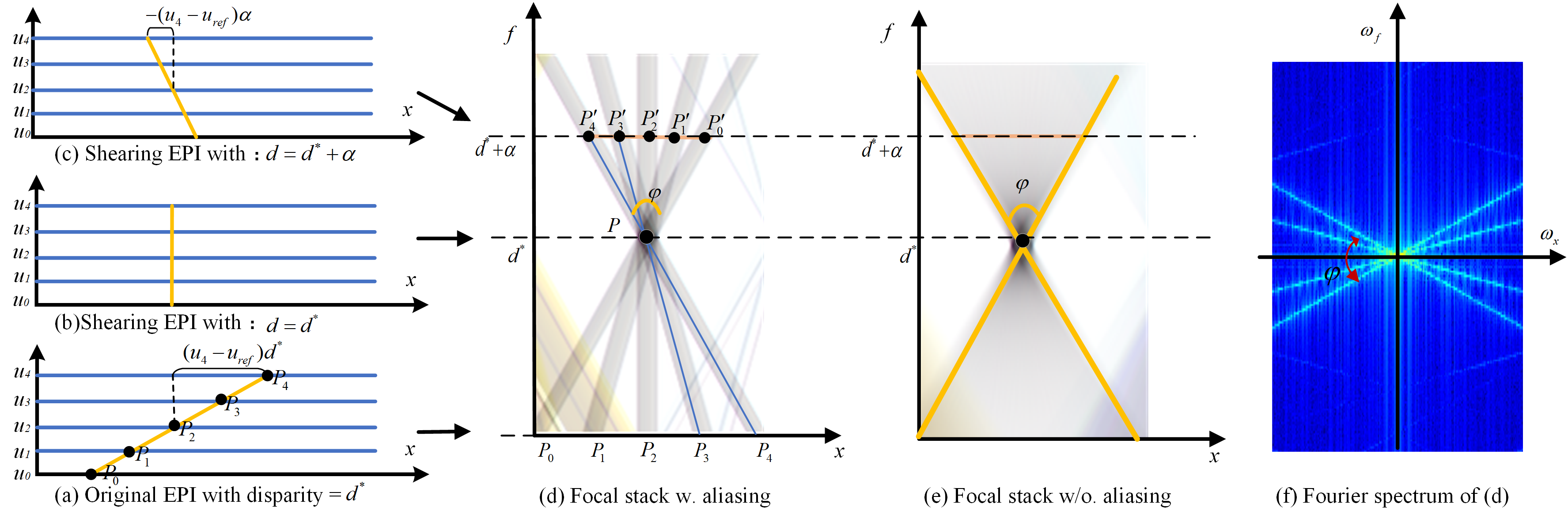}	
	
	\caption[example]
	{
	  Analysis for the EPI in the $u$-$x$ space, the focal stack in the $f$-$x$ space and the FSS in the $\omega_f$-$\omega_x$ space. (a)-(c) Original and sheared EPIs with different parameters. (d) Aliased focal stack. (e) Continuous focal stack. (f) Fourier spectrum of (d). For better visualization, only one single pixel $P$ is considered here. \label{anysis_EFS}
	}
\end{figure*}
%Due to the separability of the Fourier transform, the spectrum of a 3D focal stack could be decomposed intthis paper focuses on a 2D focal stack spectrum formed from a 2D light field to better describe the focal stack spectrum, the 2D light field representation, \textit{i.e.} EPI  is used in this section.

% \textbf{Notations}
For better understanding, the notations used in the paper are given in Tab. \ref{tab:notations}. $E(u,x)$ denotes a 2D EPI of a 4D light field, where $u$ and $x$ refer to the angular and spatial dimensions, respectively. $E_d(u,x)$ denotes the sheared EPI at a specific disparity $d$,
\begin{equation}
\small
\label{eqn:shearing_refocus}
	E_d(u,x) = E(u,x+d(u-u_{ref})),
\end{equation}
where $u_{ref}$ refers to the reference view\footnote{The central view is selected as the reference view in the paper.}. Once the sheared EPI with an arbitrary disparity $f \in [d_{min},d_{max}]$ is integrated for all views, the focal stack $F(f,x)$ is formed, 
\begin{equation}
\small
\label{eqn:focalstack_construction}
F(f,x)\!=\!\int\!\! E(u,x+f(u-u_{ref})){d}u.
\end{equation}
%=\!\int\!\! E_{f}(u,x){d}u \!

%where $f$ refers to the focal depth.
%\footnote{Noting that the disparity $d$ is equivalent to the focus parameter $f$ according to \cite{schechner2000depth}.}.

Subsequently, the Fourier form of $F(f,x)$ is %defined %$\mathcal{F}(\omega_{f},\omega_{x})$ is the
\begin{equation}
\small
\mathcal{F}(\omega_{f},\omega_{x}) = FT(F(f,x)),
\label{eqn:focal_spectrum_construction}
\end{equation}
where $FT(\cdot)$ refers to the 2D Fourier transform operator. Specifically, the view number of the given light field is $N_u$. 
%Fig.\ref{fig:define_focal_stack} demonstrates the above progress.
% need add \alpha in this scection and  Table1
\begin{table}[t]
	\small
	\begin{center}
		\caption{Notations used in the paper}
		\label{tab:notations}      % Give a unique label
		\begin{tabular}{|l|l|}
			\hline
			Term & Definition\\ \hline
			$E(u,x)$ & A 2D EPI of a 4D light field\\
			$E_d(u,x)$ & Sheared EPI at the disparity $d$  \\
			$u$ & Angular coordinate \\
			$x$ & Spatial coordinate \\
			$f$  & Focal layer's disparity\\
			$F(f,x)$ & Focal stack formed by $E(u,x)$ \\
%			$d$      & Disparity during the shearing process \\
			$\mathcal{F}(\omega_{f},\omega_{x})$ & Focal stack spectrum \\
			$FT(\cdot)$ & Fourier transform operator \\
			$u_{ref}$ &  Reference view \\
			$N_u$  & View number \\
			\hline
		\end{tabular}
	\end{center}
	
\end{table}

% \begin{tabularx}{.48\textwidth}{@{} |l|c|c|c|C|C| @{}}

%\begin{figure}[!t]
%	\begin{center}
%		\centering
%		\includegraphics[width=\linewidth]{define_focal_stack.png}
%	\end{center}
%	\caption{ The  formulation of FSS. (a) Optical path, (b) Focal stack, (c) FSS.
%}
%	\label{fig:define_focal_stack}
%\end{figure}

\subsection{Characteristics of Focal Stack and Its Spectrum}\label{sec:charFSS}	
%\sloppy{}
\label{Characteristics}
\subsubsection{Aliasing and Defocus in the Focal Stack}
\label{subsubsec:Aliasing and Defocus in the Focal Stack}
%\sloppy{}

\ 
\newline
 \indent \textbf{Sparse views and aliased focal stack.}
Fig.2(a) shows an EPI with 5 views, where the baseline between neighboring views is defined as 1 unit. A 3D point is imaged as $P_0, P_1 ,P_2, P_3, P_4$ in different views. The disparity of $P$ is $d^*$. When the EPI is sheared using Eqn.\ref{eqn:shearing_refocus} with $d=d^*$, the disparity of $P$ becomes $0$ (Fig.\ref{anysis_EFS}(b)). When the EPI is sheared with parameter $\alpha$, compared with the original EPI (Fig.\ref{anysis_EFS}(a)), the shearing parameter $d=d^*+\alpha$ and the current disparity of $P$ becomes $-\alpha$ now (Fig.\ref{anysis_EFS}(c)).

By accumulating these three EPIs along the angular axis using Eqn.\ref{eqn:focalstack_construction}, we get three focal images, \textit{i.e.}, the three dotted lines in Fig.\ref{anysis_EFS}(d). It is noticed that the distribution of $P$ varies from slice to slice. Originally, the imaging of $P$ is aliased and $P$ is decomposed as $P_0$, $P_1$, $P_2$, $P_3$, $P_4$. When the EPI is sheared with $d=d^*$, these five pixels are converged into one pixel $P$. In the third slice, $P$ is aliased again and decomposed as $P_0'$, $P_1'$, $P_2'$, $P_3'$, $P_4'$. The imaging range of $P$, \textit{i.e.}, the distance from $P_0$ to $P_4$ or $P_0'$ to $P_4'$, can be calculated by
\begin{equation}
\label{eqn:diameter_defocus}
\small
l_{P_0,P_4} = |\alpha|(N_u-1).
\end{equation}
Repeat the above shearing process with the sampling interval $\Delta\alpha$ to construct a focal stack (Fig.\ref{anysis_EFS}(d)). The distribution of $P$ forms a cone-shaped pattern in the focal stack, \textit{i.e.}, the similar triangles $\triangle  PP_0P_4$ and $\triangle PP_0'P_4'$. The apex angle $\varphi$ of such a triangle is
\begin{equation}
\label{eqn:apex_angle_dis}
\varphi = 2\arctan(\frac{1}{2}\Delta \alpha(N_u-1)).
\end{equation}
The slope of each component $PP_i$ in the $f-x$ coordinates of Fig.2(d) could be computed by
\begin{equation}
\label{eqn:slope_dis}
Slope(PP_i)=\frac{1}{\Delta \alpha(u_{i}-u_{ref})}.
\end{equation}

\noindent \textbf{Dense views and defocused focal stack.} By inserting more views between neighboring views $u_i$ and $u_{i+1}$ in Fig.2(a)(b)(c), more lines appear between the line $PP_i$ and $PP_{i+1}$ in Fig.2(d). When the views are dense enough, the aliasing becomes defocus blur (Fig.\ref{anysis_EFS}(e)). The defocus diameter of $P$ could also be calculated using Eqn.\ref{eqn:diameter_defocus}. Noting that, since the baseline between neighboring views is scaled by $\frac{1}{M+1}$ when inserting $M$ views, the defocus diameter in Fig.\ref{anysis_EFS}(e) is equal to the aliasing distance in Fig.\ref{anysis_EFS}(d). The apex angle $\varphi$ does NOT change (the same cone-shaped pattern still exists, as shown in Fig.\ref{anysis_EFS}(e)).

\noindent \textbf{Continuous focal stack.} The above analysis focuses on the focal stack with a sampling interval $\Delta \alpha$. When the focal stack is constructed in the continuous domain, Eqns.\ref{eqn:apex_angle_dis} and \ref{eqn:slope_dis} could be rewritten as
\begin{equation}
\label{eqn:apex_angle_conti}
\varphi_{con} = 2\arctan(\frac{1}{2}(N_u-1)).
\end{equation}
\begin{equation}
\label{eqn:slope_conti}
Slope(PP_i)_{con}=\frac{1}{u_{i}-u_{ref}}.
\end{equation}

According to Eqns.\ref{eqn:apex_angle_dis}-\ref{eqn:slope_conti}, we get the following observations, %{\textit{ the shape of defocus blur or aliasing line is determined by the refocus parameter $\Delta\alpha$ and the view index, and is \textbf {independent} of the scene depth. \textbf{All aliasing lines from the same view have the same slope in the focal stack.}}}

\begin{enumerate}[\indent (a)]
  \item  The shapes of defocus blur or aliasing lines are determined by the refocus parameter $\Delta\alpha$ and the view index, which is independent of depth.
  \item All aliasing lines from the same view have the same slope (as the lines $PP_i$ shown in Fig.\ref{anysis_EFS} (d)).
\end{enumerate}

\begin{figure}[t]
	\begin{center}
%		\begin{tabular}{c}			
			%\includegraphics[width=\linewidth]{deffer_samlpy_of_efs.png}
			%\subfloat[]{\includegraphics[width=1in]{deffer_samlpy_of_efs(a).png}}
			%\subfloat[]{\includegraphics[width=1in]{deffer_samlpy_of_efs(b).png}}
			%\subfloat[]{\includegraphics[width=1in]{deffer_samlpy_of_efs(c).png}}
			%\subfigure[]{\includegraphics[width=1.4in]{deffer_samlpy_of_efs(d).png}}
%		\end{tabular}
    \includegraphics[width=\linewidth]{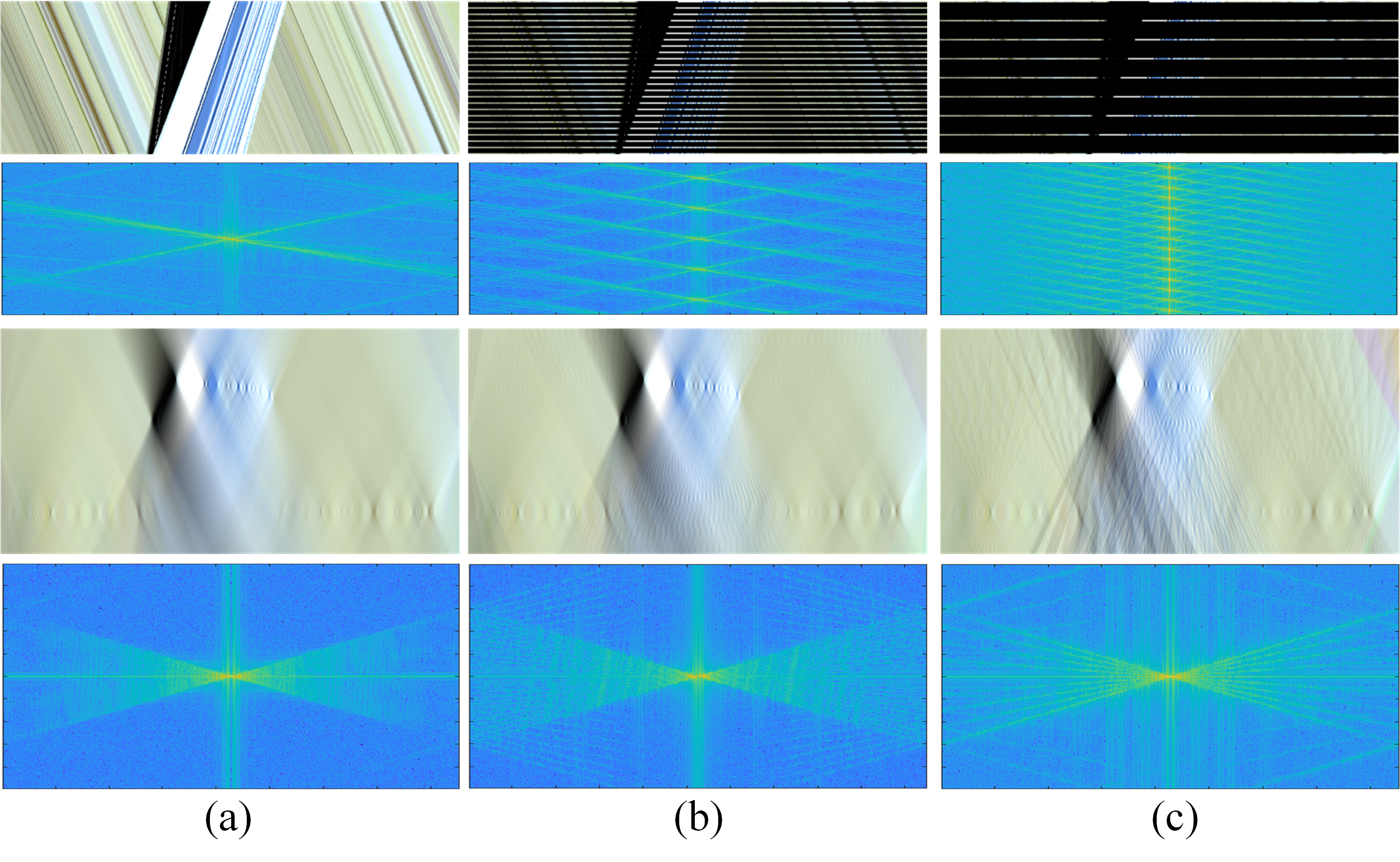}
	\end{center}
	
	\caption[example]
	{
		Comparisons of a light field under different angular sampling rates. From top to bottom: EPI and its spectrum, focal stack slice and corresponding FSS. From left to right: (a) Dense sampling with $121$ views. (b) $5\times$ downsampling. (c) $15\times$ downsampling. %(d) 30$\times$ downsampling.
		 \label{deffer_samlpy_of_efs}
	}
\end{figure}

\subsubsection{Aliasing and Defocus in the FSS}
\label{sec:Defocus and Aliasing in the FSS}

\ 
\newline
\indent Given a focal stack formed from a $N_u$-view light field, there are $N_u$ frequency lines in the FSS according to the property of the Fourier transformation \cite{bigun1987optimal} \footnote{The Fourier transformation tells that the energy of all lines with the same slope in the spatial domain will concentrate on a perpendicular line passing through the origin.}: each line corresponds to a specified view and more views lead to more lines. Consequently, the FSS also has the following properties,
\begin{enumerate}[\indent (a)]
  \item The shape of the FSS is determined by the refocus parameter $\Delta\alpha$ and the view index, which is independent of depth.
  \item According to the property of the Fourier transformation \cite{gonzales2002digital}, the FSS is conjugate symmetric. 
\end{enumerate}

% \textbf{\textit{All geometry features in the focal stack still exist in the FSS, which means the shape of the FSS is determined by the refocus parameter $\Delta\alpha$ and the view index and is independent of the scene depth.}} 
% 
%Furthermore, according to the property of the Fourier transform, the FSS is centrosymmetry at the same time. 

\subsection{EPI Spectrum vs. FSS}
\label{EPI Spectrum  VS FSS}
In this section, we will analyse the differences between the EPI spectrum and the proposed FSS. Insufficient sampling will result in repeated and overlapped aliasing patterns in the Fourier spectrum of the EPI \cite{chai2000plenoptic, mildenhall2019local, srinivasan2019pushing} and the bound on the spectral support depends on the depth range ($[z_{min},z_{max}]$). While FSS is the integral result in the frequency domain. With changing focus depths, all contents from the same view will be gathered in the same line in the FFS (refer to Sec.\ref{Characteristics}). 

Fig.\ref{deffer_samlpy_of_efs} shows the comparisons between the aliased EPI spectrum and the aliased FSS under different sampling rates. Taking a closer look at these two spectra, more repeating areas appear in the EPI spectrum with respect to the decrease in the number of views, while the structural distribution of the FSS still remains. Additionally, there is a one-to-one correspondence between the line in the FSS and the view when $\Delta\alpha$ is fixed. As shown in Fig.\ref{deffer_samlpy_of_efs}(c) (the rightmost column), 9 views in the EPI correspond to 9 lines in the FSS.

Because the EPI spectrum overlaps in the frequency domain, appropriate low-pass filtering is needed to remove the overlapping of aliased components during reconstruction or rendering. However, these low-pass filters tend to result in an over-smoothness in the focused regions. By contrast, the FSS has no overlapping spectrum so that there is no such ``spectrum isolation" problem in the proposed solution. As discussed in Sec.\ref{sec:Defocus and Aliasing in the FSS}, insufficient angular sampling causes aliasing effects in the focal stack. Considering the properties that each line of the FSS corresponds to a specific view and more views lead to more lines, the anti-aliasing problem could be formulated as a spectrum completion one based on the FSS.

\begin{figure*}[t]
\begin{center}
	\includegraphics[width=\linewidth]{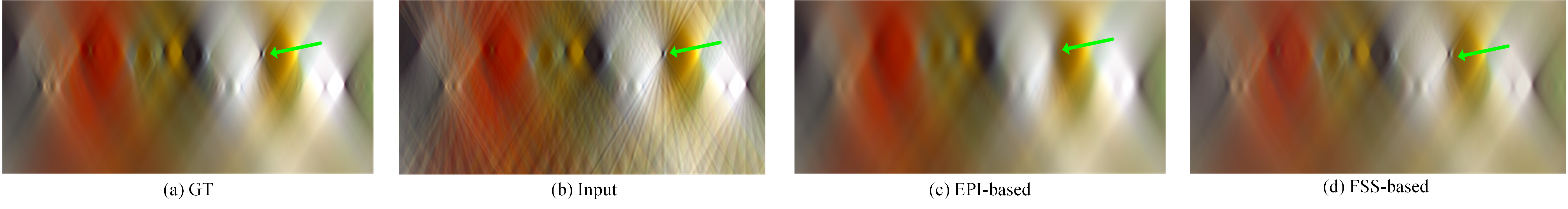}
\end{center}
	\caption[example]
	{
		$15\times$ downsampling anti-aliasing results on different focal layers. (a) GT (1$\times$ $121$ views). (b) Input (1$\times$ 9 views). Anti-aliasing results by performing (c) low-pass filtering on the EPI and (d) our FSS-based deep anti-aliasing algorithm.\label{fig:epivsfss}
	}
\end{figure*}

Fig.\ref{fig:epivsfss}(c) shows the anti-aliasing result at $15\times$ downsampling using a low-pass filter on the EPI, and Fig.\ref{fig:epivsfss}(d) shows the result using the proposed FSS-based deep anti-aliasing algorithm. The EPI-based result is over-smoothed at the focused points (the area pointed by the green arrow). The proposed method not only eliminates the aliasing effects but also maintains the sharpness of focused points and edges.

\begin{figure*}[t]
\begin{center}
	\includegraphics[width=\linewidth]{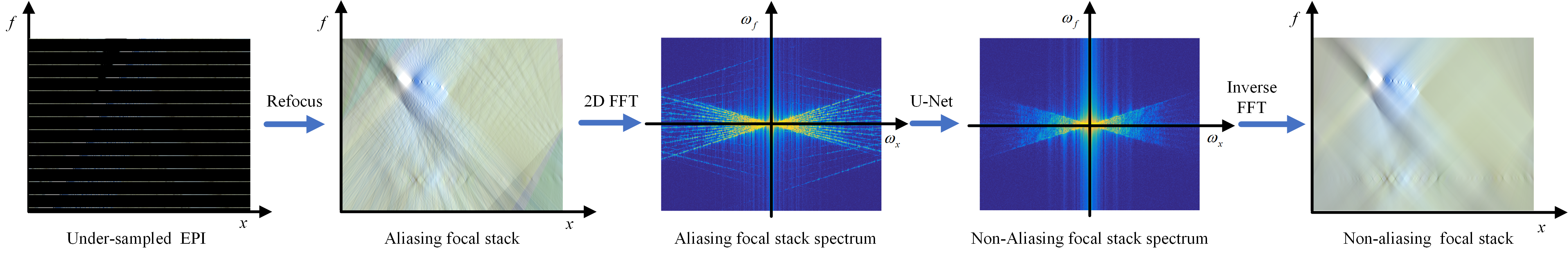}
\end{center}
	\caption[example]
	{
		The pipeline of the proposed FSS-based deep anti-aliasing algorithm. \label{pipeline}
	}
\end{figure*}

\section{Proposed Method}
\sloppy{}

As mentioned in Sec. \ref{sec:charFSS}, aliasing effects are caused by insufficient angular sampling and there is a relationship between the number of views and the FSS. As a result, the anti-aliasing problem could be modeled as a spectrum completion one. The pipeline of the proposed FSS-based anti-aliasing algorithm is shown in Fig.\ref{pipeline}.

% so we proposed an anti-aliasing method based on focal stack spectrum $\mathcal{F}$  reconstruction as a spectrum completion problem. By utilizing the properties of focal stack spectrum in the Fourier domains, we can restore the spectrum without aliasing by learning the spectrum bands corresponding to missing viewpoints by undersampling.

Specifically, the aliasing focal stack $F_a(f,x)$ is obtained from the undersampled EPI using Eqn.\ref{eqn:focalstack_construction}. Then the Fourier transform operator $FT(\cdot )$ is applied to obtain the FSS $\mathcal{F}_{a}(\omega_{f},\omega_{x})$. Finally, a CNN $\phi$ parameterized by $\sigma$ is proposed to reconstruct the aliasing-removed FSS $\mathcal{F}(\omega_f,\omega_x)$ from $\mathcal{F}_a(\omega_f,\omega_x)$, % via the follow optimization

\begin{equation}
%\small
  \label{EFS_reconstruct}
\underset{\sigma}{\arg \min }\left\{
\left\|\mathcal{F}_{gt}, \phi_{\sigma}(\mathcal{F}_a)\right\|
\right\},
\end{equation}
where $\mathcal{F}_{gt}$ is the ground truth FSS.

\begin{figure}
	\begin{center}			\includegraphics[width=\linewidth]{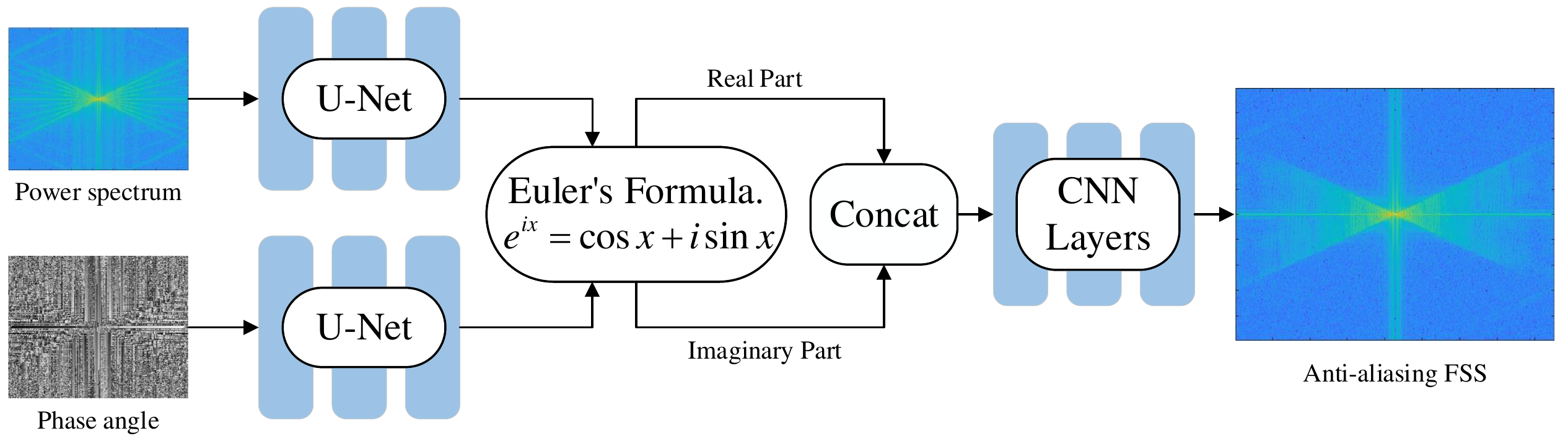}
	\end{center}
	\caption
	{
		The structure of the proposed FSS completion network. \label{fig:network}
	}
\end{figure}

\begin{figure}
\begin{center}
	\begin{tabular}{c}
		\includegraphics[width=\linewidth]{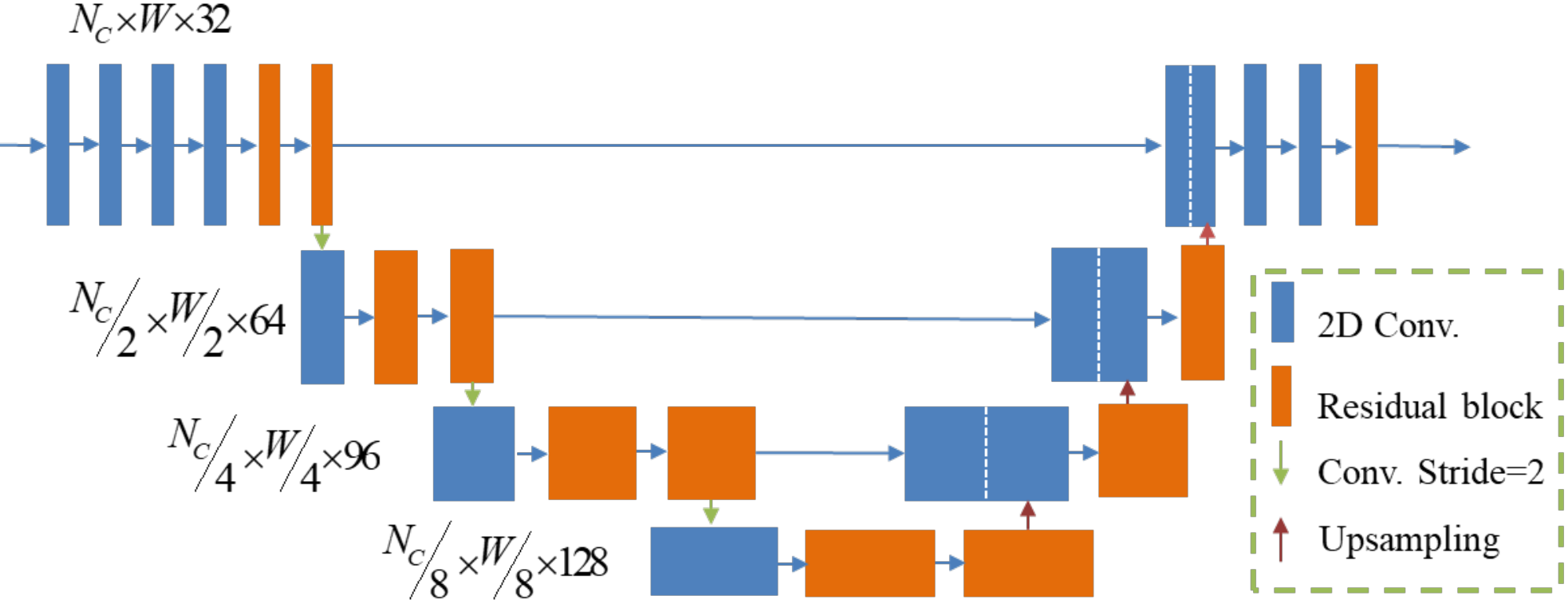}
	\end{tabular}
\end{center}
	\caption[example]
	{
		The U-Net architecture adopted in the paper \cite{ronneberger2015u}. 	
	}
\label{network1}
\end{figure}

As shown in Fig.\ref{fig:network}, to deal with complex number inputs, a dual-stream U-Net \cite{ronneberger2015u} is designed. The power spectrum and the phase angle are firstly fed into two sub-networks respectively. Then the generated features are combined using the Euler's formula to obtain the real and imaginary parts, which are concatenated and fed to another network for optimization. Fig.\ref{network1} shows more details of the adopted U-Net architecture.

%Fig.\ref{network} shows the network architecture.

The loss function is,
\begin{equation}  \label{loss}
%\small
{ loss }=\Vert\mathcal{F}-{\mathcal{F}_{g t}}\Vert_{2}+\lambda\cdot{loss}_{s},
\end{equation}
where ${loss}_{s}$ constrains the conjugate symmetry of the reconstructed FSS. The scalar $\lambda$ is set to 1.5 for balancing the two loss terms,
% $\lambda=1.5$ balances the weights of two loss terms,

%{loss}_{r}=\sqrt{\frac{1}{CW} \sum\limits_{i=0}^{C-1} \sum\limits_{j=0}^{W-1}|\mathcal{F}(\omega_{i}, \omega_{j})-{\mathcal{F}_{g t}(\omega_{i}, \omega_{j})}|^2},

\begin{equation} \label{loss:details}
 \small
{loss}_{s}=\frac{1}{N_{C} W} \sum\limits_{i=0}^{N_{C}-1} \sum\limits_{j=0}^{W-1}\left| \mathcal{F}(\omega_{i}, \omega_{j})-  {\mathcal{F}^*(-\omega_{i}, -\omega_{j})} \right|,
\end{equation}
%&{loss}_{s}=\frac{1}{C W} \sum\limits_{i=0}^{C-1} \sum\limits_{j=0}^{W-1}\left| |\mathcal{F}(\omega_{i}, \omega_{j})|-  |{\mathcal{F}(-\omega_{i}, -\omega_{j})}| \right|,
where  $|\cdot|$ refers to the norm of a complex number, $*$ indicates the standard conjugate operation on a complex number. $N_C$ and $W$ denote the number of refocus layers and the image width respectively. 

The complete FSS-based deep anti-aliasing algorithm is summarized in Algorithm~\ref{algorithm1}. %Given a 3D light field having $N_u\times$ views and $H$ and $W$ are the image height and width.
% the horizontal EPI of each line is sequentially extracted for constructing an aliasing 2D focal stack and the corresponding spectrum, then the network mentioned above is used for completing the

\begin{algorithm}
	\caption{The FSS-based deep anti-aliasing algorithm}
	\label{algorithm1}
	\begin{algorithmic}[1]
		\renewcommand{\algorithmicrequire}{\textbf{Input:}}
		\renewcommand{\algorithmicensure}{\textbf{Output:}}
		\REQUIRE ~~\\ An angularly undersampled 3D light field $L(u,x,y)$, with $N_u$ views of $H \times W$ pixels.		
		\ENSURE ~~\\ %Output
		An anti-aliasing 3D focal stack.
		\FOR {$y=1$ to $H$}
		\STATE  Get the 2D light field $ E(u,x)$.\\
%		\STATE  Perform the shearing operation on $E(u,x)$ by Eqn.\ref{eqn:shearing_refocus}.\\
		\STATE  Obtain the focal stack slice $F_a(f,x)$ by Eqn.\ref{eqn:focalstack_construction}. \\
		\STATE  Get the aliasing FSS $\mathcal{F}_{a}(\omega_{f},\omega_{x})$ by Eqn.\ref{eqn:focal_spectrum_construction}.\\
		\STATE  Reconstruct the aliasing-reduced FSS $\mathcal{F}(\omega_{f},\omega_{x})$ with the dual-stream U-Net.
		\STATE  Perform an inverse Fourier transform on $\mathcal{F}(\omega_{f},\omega_{x})$.
		
		\ENDFOR
		\STATE Get the anti-aliasing 3D focal stack.
	\end{algorithmic}
%	\vspace{-1.5em}
\end{algorithm}

\section{Experimental Results}
\label{Experimental Results}
\sloppy{}
In this section, we demonstrate the performance of the proposed FSS-based anti-aliasing algorithm on both synthetic and real light fields. The robustness of the proposed method is firstly validated on light fields with different sampling rates. Then, an ablation experiment is conducted to verify the effectiveness of the conjugate-symmetric loss. Finally, both the quantitative and qualitative comparisons with SOTAs are provided to demonstrate the superiority of the proposed method. Additional experiments on light fields captured from a camera array further verify the generalization of the proposed method. 

\subsection{Datasets and Implementation Details}
\sloppy{}
%We use the real light field dataset proposed in Guo $et~al.$\cite{guo2018dense}. In order to show the relationship between views and FSS lines,  the first 121 views are used. This dataset contains a large number of real static scenes, such as bicycles, toys and plants, which have abundant colors and complicated occlusions. We also render six synthetic light fields using POV-Ray \cite{povray_web} which contains prominent differences in the contents. The number of angular resolution is also $1\times 121$. The spatial resolution of real scenes and synthetic scenes are $376\times526 $ and $526\times526$ respectively.
%
%To train and verify the proposed network, both the synthetic and real light fields are used. For synthetic data, 6 light fields(LF) with $1\times 121\times 526\times 526$ are rendered using the POV-Ray\cite{povray_web}, where 4 for training and 2 for testing. For real data, the high-resolution light field dataset captured by Guo ${\it et al.}$.\cite{guo2018dense} are used, where 10 for training and 2 for testing, noting that, only the 121 views are used. Please refer the  supplementary material for the selection of light fields.

To train and verify the proposed network, both synthetic and real light fields are used. For synthetic data, 6 light fields are rendered using the light field automatic generator \cite{zhu2019revisit} and POV-Ray \cite{povray_web}, of which 4 for training and 2 for testing. For real data, the high angular resolution light field datasets \cite{guo2018dense} are used, of which 10 for training and 2 for testing. Note that only $121$ views are used. Additionally, the Stanford \cite{stanford_data} and the Disney \cite{kim2013scene} light fields are used to verify the performance of the proposed method on unseen light fields captured by a camera array. Tab. \ref{tab:data parameters} shows the details of these light field datasets. Notice that, the spatial resolutions for Couch and Church light fields ($2670\times4020$ and $2622\times 4007$ respectively) are resized in our experiments. 

\begin{figure}
\begin{center}
\includegraphics[width=\linewidth]{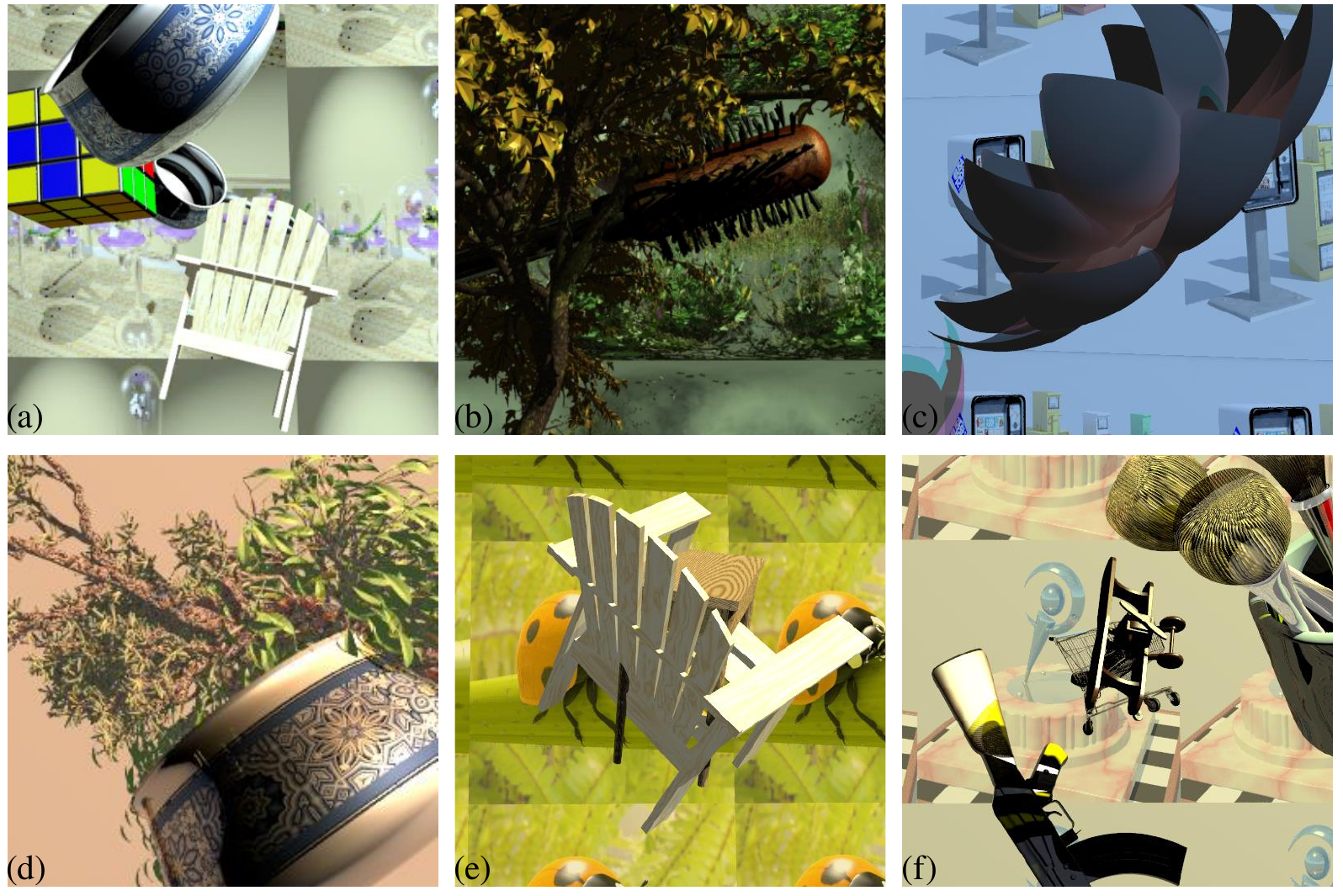}
\end{center}
\caption[example]
{
	Reference views of our synthetic light field datasets.
}
\label{view}
\end{figure}

\noindent \textbf{Details of light field scene selection.}
Fig.\ref{view} shows the reference views in our synthetic light field datasets. In the synthetic datasets, the scenes Pot-cube (Fig.\ref{view}(a)) and Tree (Fig.\ref{view}(b)) are used for testing, since the occlusions in these two scenes are more complex. In the real light field datasets \cite{guo2018dense}, the scenes Bicycle and Hydrant are selected for testing due to the larger disparity ranges, which could better verify the generalization of the proposed method under different scenes. In the Disney datasets \cite{kim2013scene}, the scenes Couch and Church are selected for testing since they are motion-free and exhibit larger disparities. Besides, we also choose the StillLife scene from the HCI datasets \cite{hci_web} for testing. Different from other datasets which provide sufficient views, the angular sampling of StillLife is inadequate, \textit{i.e.}, there is aliasing in refocused images even all the views are used. The proposed method is capable of eliminating the implied aliasing effect in this scene, as shown in Fig.\ref{fig:anti_aliasing_hci_data}.

%In order to verify the capability of the proposed method for large disparities area, we conduct experiments with different angular sampling rates, \textit{i.e.}, $5\times$ and $15\times$ downsampling scales. Please refer the SM for the sampling patterns. 

\begin{figure}
	\begin{center}
	\includegraphics[width=\linewidth]{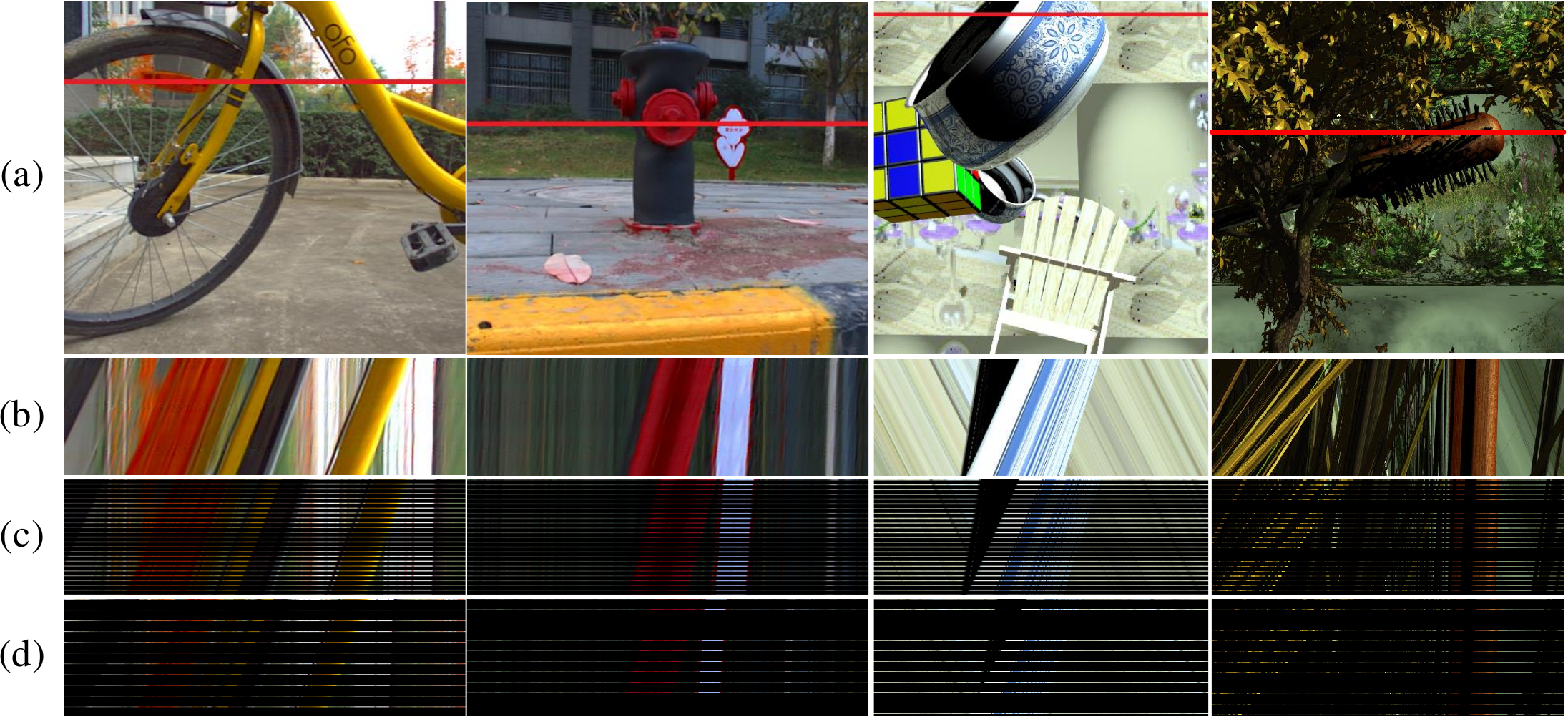}
		
	\end{center}
	\vspace{1.5em}
	\caption[example]
	{
		EPIs under different view downsampling rates. (a) Reference view. (b) Original EPI. (c) $5\times$ downsampling. (d) $15\times$ downsampling.
	}
	\label{differ_epi_samply}
\end{figure}

In order to verify the capability of the proposed method for large disparity, we conduct experiments with different angular sampling rates. At present, only single direction disparity is concerned, so the 2D EPI can be used to represent the input light field. Fig.\ref{differ_epi_samply} shows details of testing light fields under $5\times$ and $15\times$ downsampling scales. For each light field, the focal stack $F(f,x)$ is constructed by performing refocusing operations 199 times with $\Delta\alpha=0.01$. Tab. \ref{tab:data parameters} shows the ranges of refocus operation ($d$). %Fig.\ref{differ_epi_samply} shows the details of tested light field under 5$\times$ and 15 $\times$ downsampling rates. 

%Fig.\ref{differ_epi_samply} shows the tested light fields under $5$ and $15$ downsampling scales.

%For each light field in this paper, we performed 199 refocusing operations and the value of $\Delta\alpha$ is 0.01 in Eqn.\ref{eqn:slope_aliasing_line}. Tab.\ref{tab:data parameters} shows the detail of experimental and data parameters of different light field. Notice that, the spatial resolution of couch and church light field are resized in ours experimental.
%,  so we can deal with the aliasing effects from different focal depths simultaneously and maintain the consistency of the refocused image along the focal direction(PSF-continuity).

The network converges after 150 epochs where each epoch contains 30 iterations. The Adam optimizer \cite{kingma2014adam} is used for iterative optimization. The learning rate is initially set to 0.001. %$1e\!-3$. 
The first and second moments of the gradients are set to 0.9 and 0.99 respectively to enable adaptive learning rates. The network is implemented using the TensorFlow framework with 7 GTX 1080Ti GPUs.

\begin{table}[t]
\footnotesize
\begin{center}
\caption{ Details of light fields evaluated in the experiments.}
\label{tab:data parameters}
\begin{tabular} {lccc}
\hline 
Dataset & Rang of $d$ & Angular Res. & Spatial Res.\\
\hline
Syn. LF & [-1.00,0.98] & $ 1\times 121 $ & $526\times 526$\\ %\hline
Real LF \cite{guo2018dense} &[-1.00,0.98] & $ 1\times 121 $ & $376\times 526$\\ %\hline
Couch \cite{kim2013scene}& [-2.60,-0.60] & $ 1\times 101 $ &  $628\times 1024 $\\ %\hline
Church \cite{kim2013scene}&[-1.45,-0.53] & $1\times 101 $ &  $670\times 1024$\\ %\hline
%StillLife\cite{hci_web}&[-1,0.98] & $1\times 9$  & $768\times 768$\\
%\hline
Lego \cite{stanford_data}&[-1.00,0.98] & $1\times 17 $ & $1024\times 1024$\\ %\hline
\hline 
\end{tabular}
\end{center}
\end{table}

\begin{figure*}
	\begin{center}			\includegraphics[width=\linewidth]{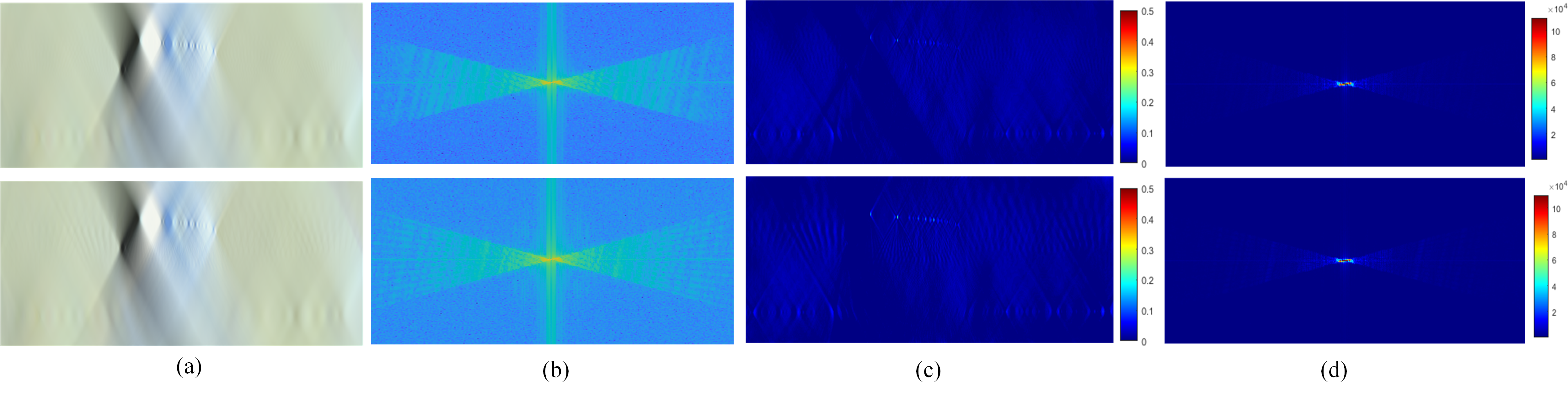}	
	\end{center}
	
	\caption[example]
	{
		Reconstructed focal stack slices and FSS under different downsampling settings. Top: $5\times$ downsampling. Bottom: $15\times$ downsampling. (a) Anti-aliasing on the focal stack. (b) Reconstructed FSS. (c) Error map of the focal stack. (d) Error map of the FSS.
	}
	\label{focalstack_errormap}
\end{figure*}

\begin{figure*}[!]
	\begin{center}
	\includegraphics[width=\linewidth]{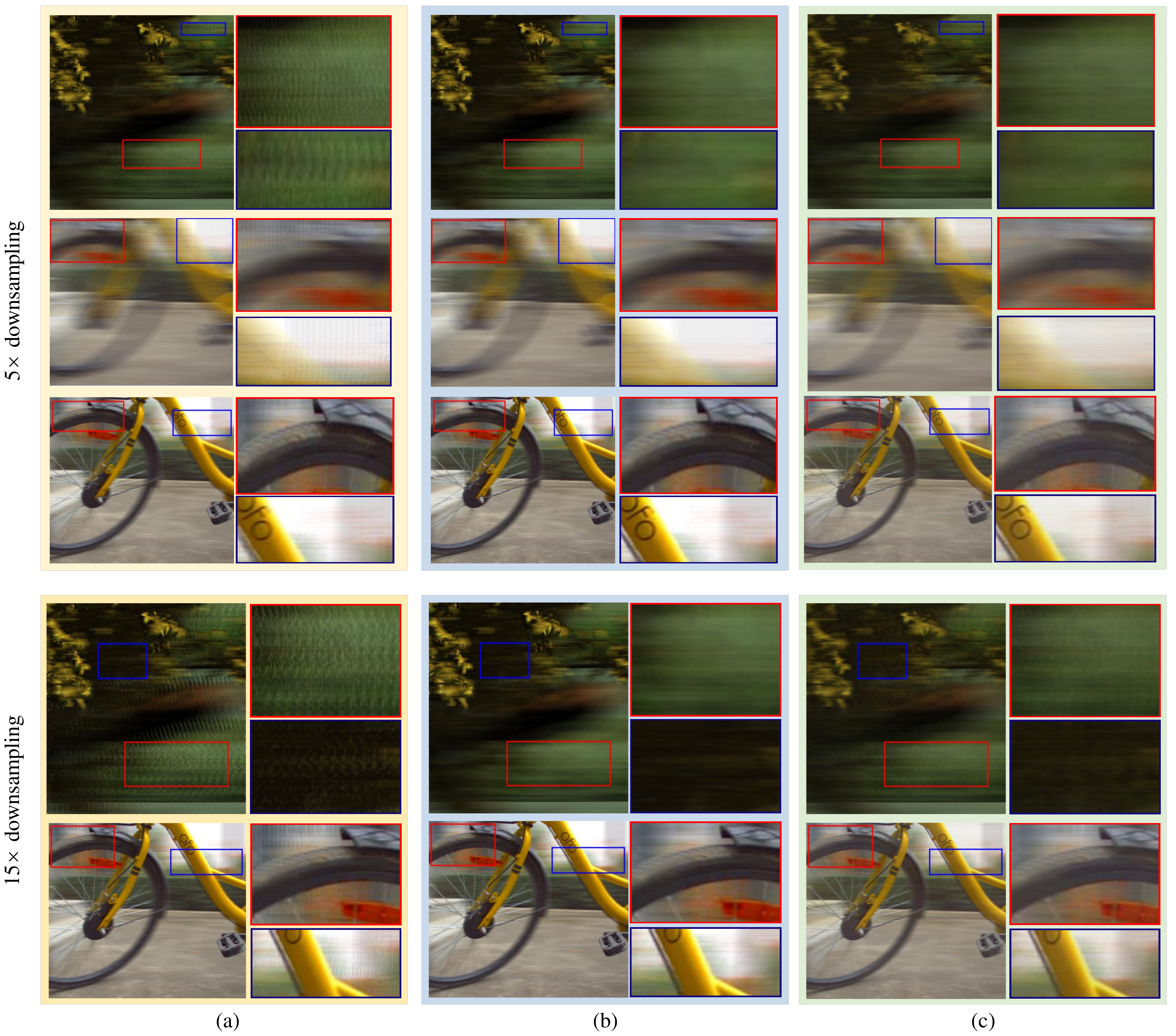}
	\end{center}
%\vspace{-5pt}
	\caption[example]
	{
		Aliasing-removed results at different focal layers under $5\times$ and $15\times$ downsampling rates. (a) Input image. (b) Ground truth. (c) Our result.
	}
	\label{self-construct-05-15}
\end{figure*}

\subsection{Method Capacity}
\label{sec:Method Capacity}
\sloppy{}
% \begin{table}[t]
% 	\footnotesize
% 	\begin{center}
% 		\caption{Average spectral energy loss on different test scenarios.}
% 		\label{tabspectral energy loss}
% \begin{tabular}{  p{2cm}<{\centering}|p{2cm}<{\centering}|p{2cm}<{\centering} }
% \hline \hline
% Dataset & 5$\times$  & 15$\times$  \\
% \hline
% Syn. LFs &2.03\% & 3.58\%\\ \hline
% Real LFs \cite{guo2018dense} &2.50\%& 3.13\%\\ \hline
% Lego \cite{stanford_data}&  \multicolumn{2}{c}{0.56\% (2$\times$ )} \\
% \hline
% Couch \cite{kim2013scene}&  \multicolumn{2}{c}{2.71\% (10$\times$ )} \\
% \hline
% Church \cite{kim2013scene}&  \multicolumn{2}{c}{0.67\% (10$\times$ )} \\
% \hline \hline
% \end{tabular}
% 	\end{center}
% 	\vspace{-2.0em}
% \end{table}

\begin{table}[t]
	\footnotesize
	\begin{center}
		\caption{Average spectral energy loss on different test scenarios.}
		\label{tabspectral energy loss}
\begin{tabular}{  p{2cm}<{\centering}|p{2cm}<{\centering}|p{2cm}<{\centering} }
\hline
Dataset & $5\times$  & $15\times$  \\
\hline
Syn. LFs &2.03\% & 3.58\%\\ \hline
Real LFs \cite{guo2018dense} &2.50\%& 3.13\%\\ \hline
Lego \cite{stanford_data}&  \multicolumn{2}{c}{0.56\% ($2\times$)} \\
\hline
Couch \cite{kim2013scene}&  \multicolumn{2}{c}{2.71\% ($10\times$)} \\
\hline
Church \cite{kim2013scene}&  \multicolumn{2}{c}{0.67\% ($10\times$)} \\
\hline
\end{tabular}
	\end{center}
	\vspace{-2.0em}
\end{table}

\begin{figure*}
	\begin{center}			\includegraphics[width=\linewidth]{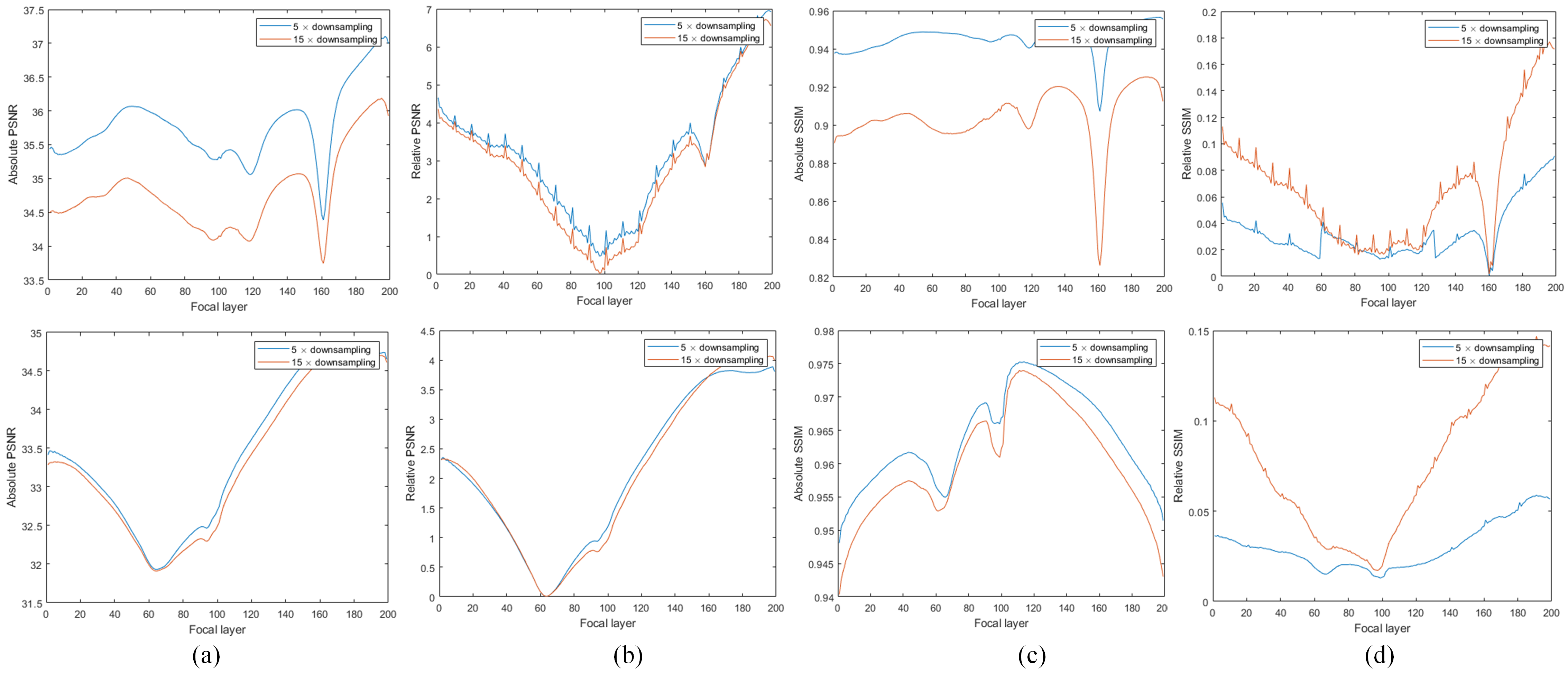}
		
	\end{center}

	\caption[example]
	{
		Quantitative results of our method across focal layers under different downsampling settings. Top: Tree scene. Bottom: Bicycle scene. (a) Absolute PSNR. (b) Relative PSNR. (c) Absolute SSIM. (d) Relative SSIM. \label{self-psnr-ssim}
	}
\end{figure*}

%The proposed method can eliminate different degrees of aliasing via one unified solution. 

\noindent \textbf{Spectrum Domain}. In this subsection, we demonstrate the performance of our approach with respect to different sampling rates. Comparing Fig.\ref{deffer_samlpy_of_efs}(b)(c) with Fig.\ref{focalstack_errormap}, we can see that, for different sampling rates, our method achieves a preferable performance on anti-aliasing rendering in both spatial and frequency domains. It is important to note that the errors in the frequency domain mainly come from the Direct Component (DC) of the spectrum. The main reason is that the energies of DC and its near neighborhoods are much larger than those of other components \cite{gonzales2002digital}. Also deep learning is more inclined to learn low-frequency components from an input signal map \cite{rahaman2019spectral}. The differences will cause color defects during the FSS reconstruction stage. To tackle this problem, we replace the DC and its near neighborhoods ($5\times5$ patch around the central pixel) of the output spectrum with those of the input spectrum. Tab. \ref{tabspectral energy loss} shows the average spectral energy loss on different test scenarios. From Tab. \ref{tabspectral energy loss} and  Tab. \ref{tab:average of quantitative results}, it is found that the PSNR of our method is greatly affected by energy loss, while SSIM is little affected by it. According to \cite{ssim_paper}, human visual system is more sensitive to the structural similarity than PSNR. Based on this assumption, the errors of defocus blur in refocused images are hard to detect by human eyes. Therefore a higher SSIM value in Tab. \ref{tab:average of quantitative results} demonstrates better performance of the proposed method.

\noindent \textbf{Image Domain}. Fig.\ref{self-construct-05-15} shows the results of anti-aliasing in different focal images under $5\times$ and $15\times$ downsampling settings. For the $5\times$ downsampling setting, the proposed method could remove aliasing around severely occluded objects (trees in the top row). The middle and bottom rows show the anti-aliasing results at different focal layers for the same scene (Bicycle) when the focused depth is beyond and within the range for the scene depth respectively. Although the radius of defocus blur varies, the proposed method could simultaneously eliminate the aliasing effects (the middle row) and retains the sharp edges (the bottom row), which shows the PSF-continuity is well maintained by the proposed method. For the $15\times$ downsampling setting, although the aliasing is more severe due to the larger disparity, our method can still obtain satisfied aliasing-removed results.

% The middle and bottom rows show the anti-aliasing results at different focal layers for the same scene (Bicycle). The middle and bottom rows show the results for the Bicycle scene when the focused depth is out of the range of the scene depth and in the range of  scene depth, respectively. 
%
%

%
%\begin{figure}
%	\begin{center}
%		\begin{tabular}{c}
%			\includegraphics[width=\linewidth]{self-construct-05.pdf}
%		\end{tabular}
%	\end{center}
%	\caption[example]
%	{
%		Aliasing removing results at different focal layers under 5$\times$ downsampling. (a) Input image. (b) Ground truth. (c) Our result.\label{self-construct-05}
%	}
%\end{figure}
%
%
%\begin{figure}
%	\begin{center}
%		\begin{tabular}{c}
%			\includegraphics[width=\linewidth]{self-construct-15.pdf}
%		\end{tabular}
%	\end{center}
%	\caption[example]
%	{
%		Aliasing removing results at different focal layers under 15$\times$ downsampling. (a) Input image. (b) Ground truth. (c) Our result.\label{self-construct-15}
%	}
%\end{figure}

%We conduct experiment evaluations using two established metrics, Peak Signal to Noise Ratio (PSNR) and Structural Similarity
%Index Measure (SSIM).
Fig.\ref{self-psnr-ssim} shows the quantitative results of our method across focal layers under different downsampling settings. The relative PSNR is obtained by subtracting the PSNR of the input focal layer image from the absolute PSNR. The same operation holds for the relative SSIM. As shown in Fig.\ref{self-psnr-ssim}(b)(d), both PSNR and SSIM fluctuate with refocusing depth. {It is noticed that, the relative PSNR/SSIM curves are not smooth. We believe it is due to the Tree scene is composed of trees with complex occlusion (see rightmost column of Fig.\ref{differ_epi_samply}), which causes many jitters in the PSNR/SSIM curves of the input focal stacks.} Also when there is mild aliasing in the refocused image, the promotion of PSNR and SSIM is less obvious (see the bottom row in Fig.\ref{self-construct-05-15}). However, the varying trends of PSNR/SSIM curves demonstrate our approach maintains the continuity of PSF in the focal stack.

\begin{figure}
\footnotesize
	\begin{center}			\includegraphics[width=\linewidth]{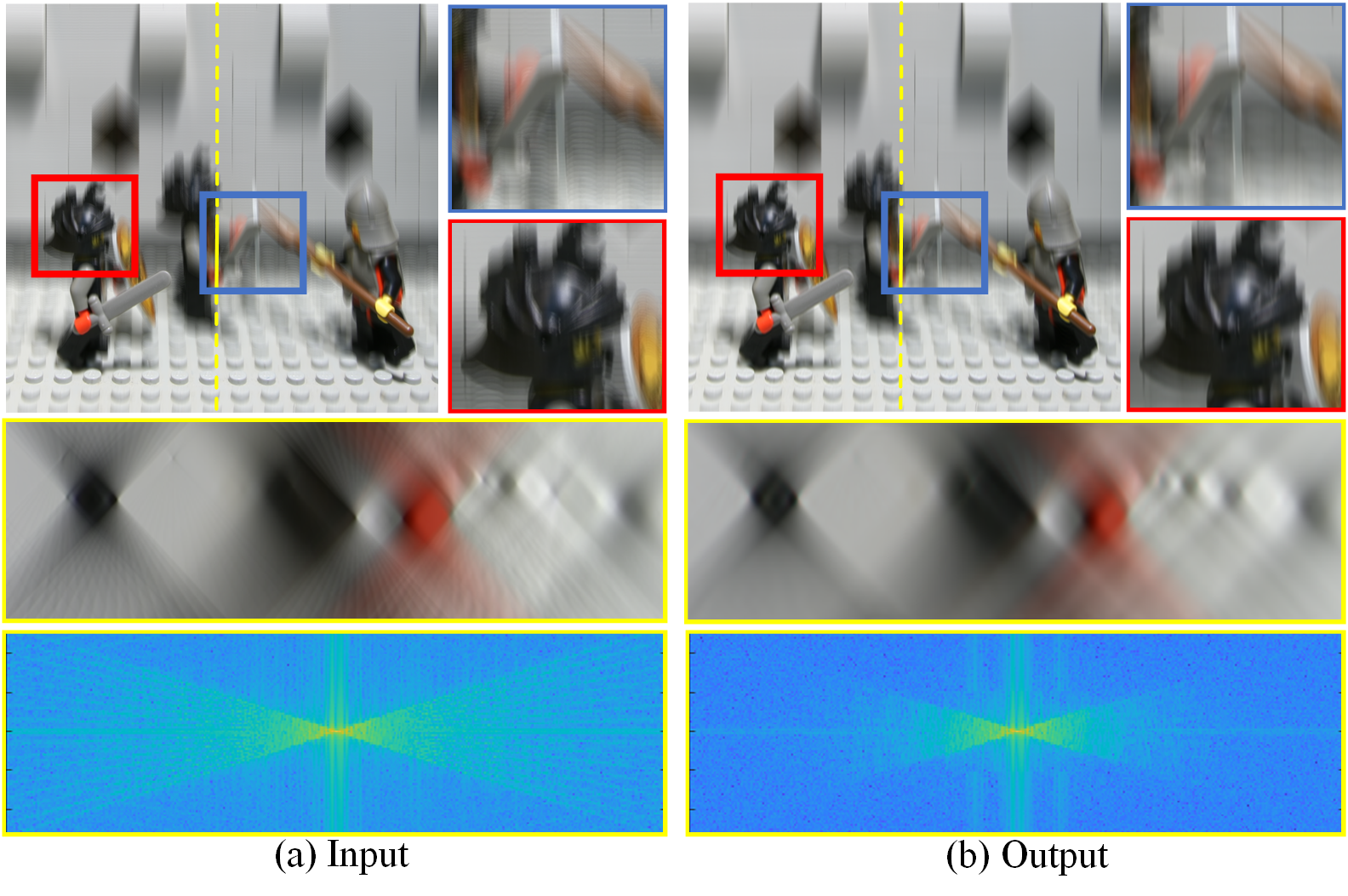}
\end{center}
\caption[example]
{
	Anti-aliasing results on the Lego Knights scene \cite{stanford_data} along the directions of $y-v$. From top to bottom: refocused image at a certain depth, partial focal stack along the yellow solid line, and corresponding FSS. (a) Input with aliasing (the view number is $17\times 1$); (b) Anti-aliasing result. \label{fig:anti-aliasling_along_yv}
}
\end{figure}

\begin{figure}
\begin{center}
	\includegraphics[width=\linewidth]{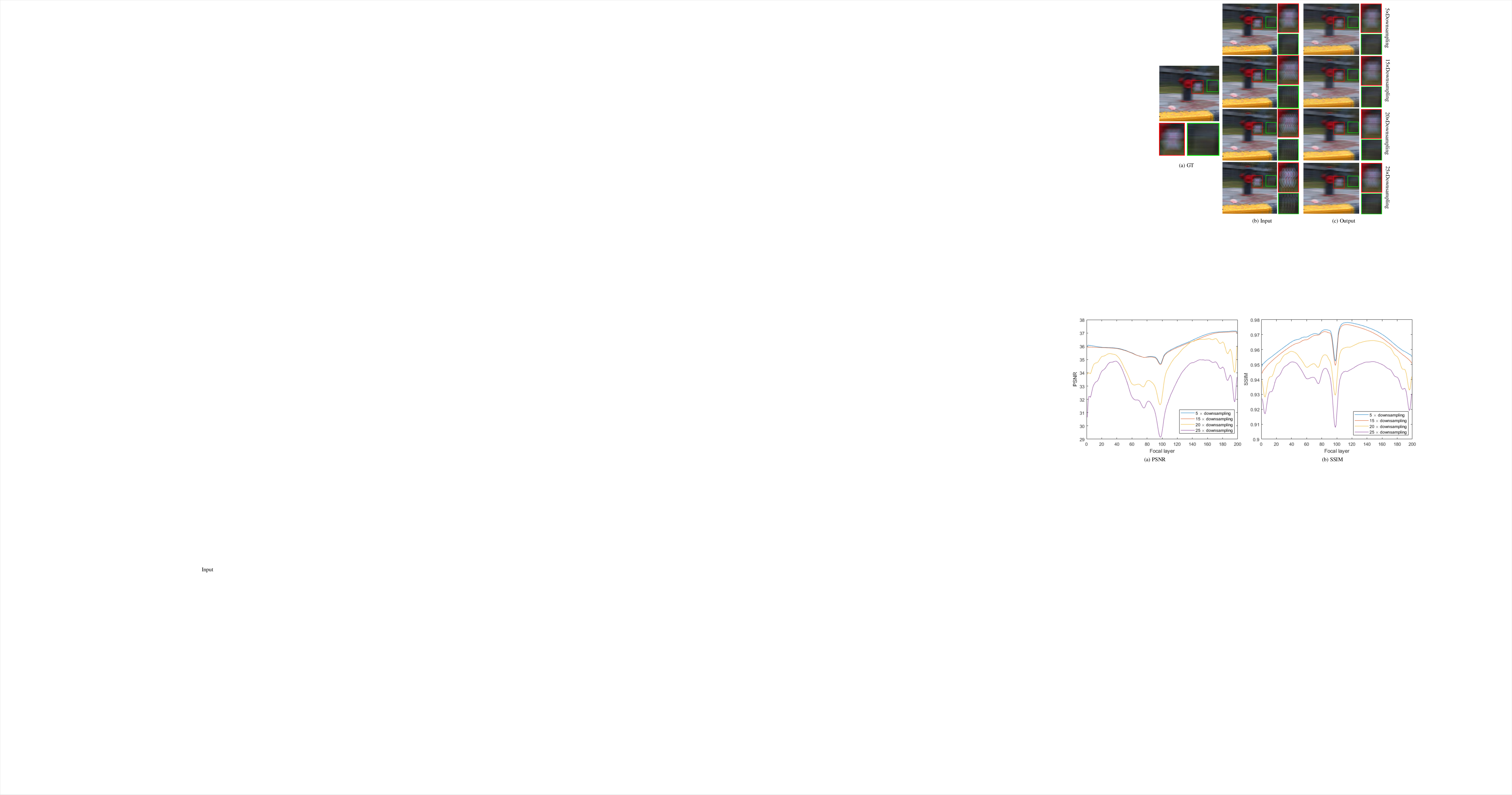}
\end{center}
\caption[example]
{
	Aliasing-removed results under $5\times$, $15\times$, $20\times$ and $25\times$ downsampling rates. (a) Ground truth. (b) Input. (c) Output.
}
\label{fig:tolerance}
\end{figure}

\begin{figure}
\begin{center}
	\includegraphics[width=\linewidth]{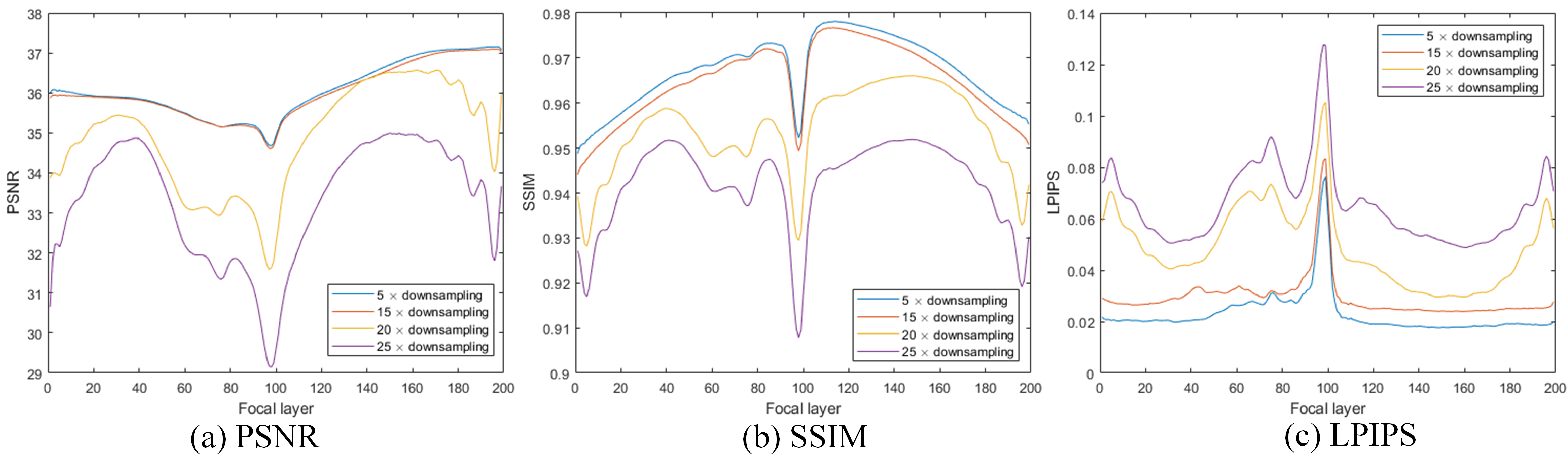}
\end{center}
\caption[example]
{
	Quantitative comparisons of anti-aliasing results in Fig.\ref{fig:tolerance}. %The blue, red and yellow lines indicate the results by ours, Kalantari{\it et al.} \cite{kalantari2016learning} and Xiao {\it et al.} \cite{xiao2017aliasing} respectively. 
% The blue lines are ours results, the red lines are  Kalantari {\it et al.} \cite{kalantari2016learning} results and the yellow lines are obtain by Xiao {\it et al.} \cite{xiao2017aliasing}.
}
\label{fig:tolerance_psnr_ssim}
\end{figure}

\begin{figure}
\begin{center}

	\includegraphics[width=\linewidth]{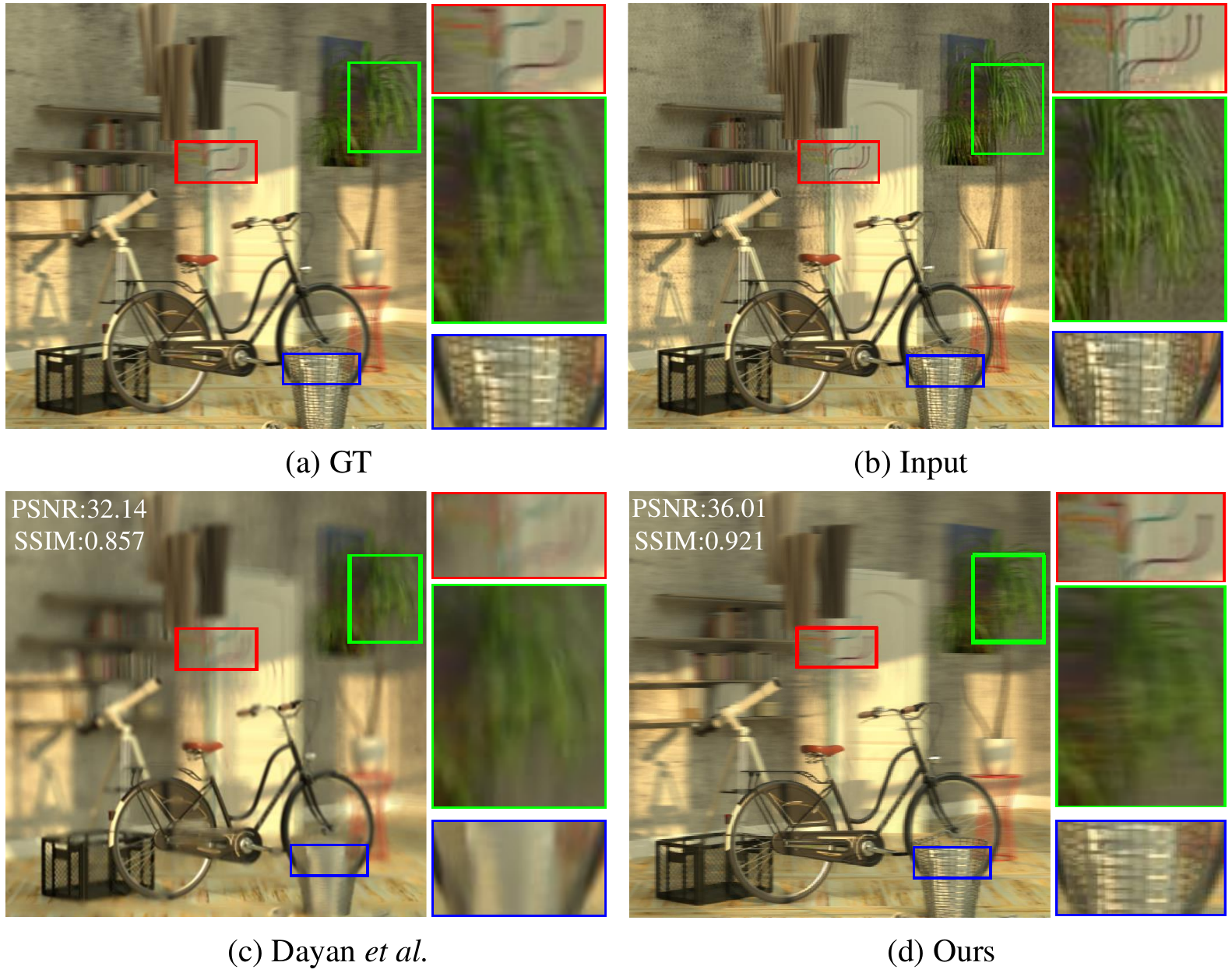}

\end{center}
\caption[example]
{
	Anti-aliasing results on Bicycle from the HCI datasets \cite{honauer2016dataset} when refocusing at $d=-0.75$. (a) Ground truth. (b) Input with aliasing (the number of views is $1\times2$). Results by (c) Dayan {\it et al.} \cite{ben2020deep} and (d) our method. 
}
\label{fig:Minimal angular}
\end{figure}

\begin{figure*}[!]
\begin{center}
	\includegraphics[width=1.00\linewidth]{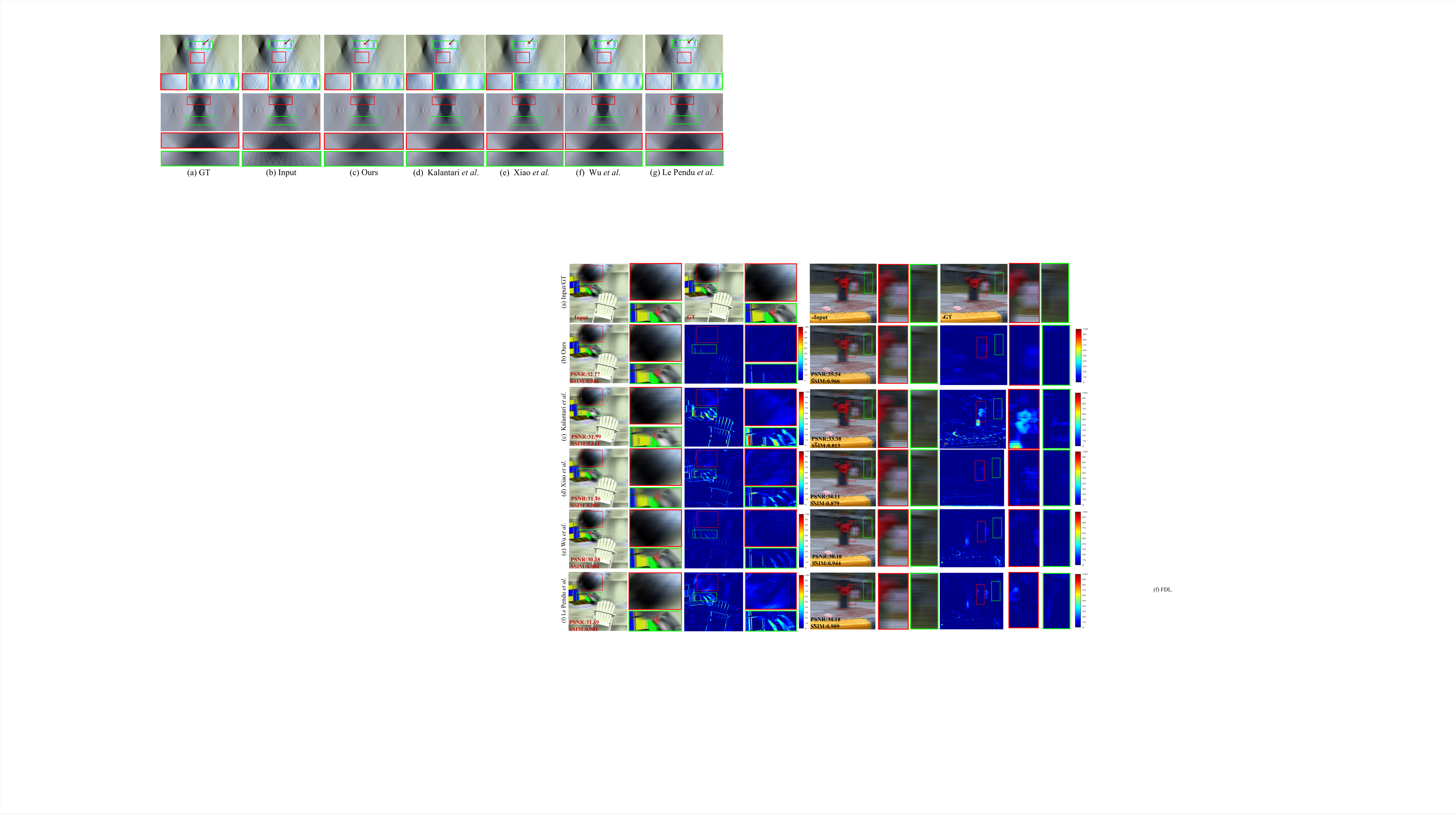}
\end{center}
\vspace{-1.0em}
\caption[example]
{
	Anti-aliasing results under $15\times$ downsampling (refocused images and error maps). (a) Input focal image and ground truth. Results by (b) our method, (c) Kalantari {\it et al.} \cite{kalantari2016learning}, (d) Xiao {\it et al.} \cite{xiao2017aliasing}, (e) Wu {\it et al.} \cite{wu2017light_epi} and (f) Le Pendu {\it et al.}  \cite{pendu2019fourier}. The left half shows synthetic LFs and the right half shows real LFs. Several local areas are zoomed in for better visualization. \label{contrast_3_15_72}
}
\end{figure*}

\begin{figure*}
\begin{center}
	\includegraphics[width=1.00\linewidth]{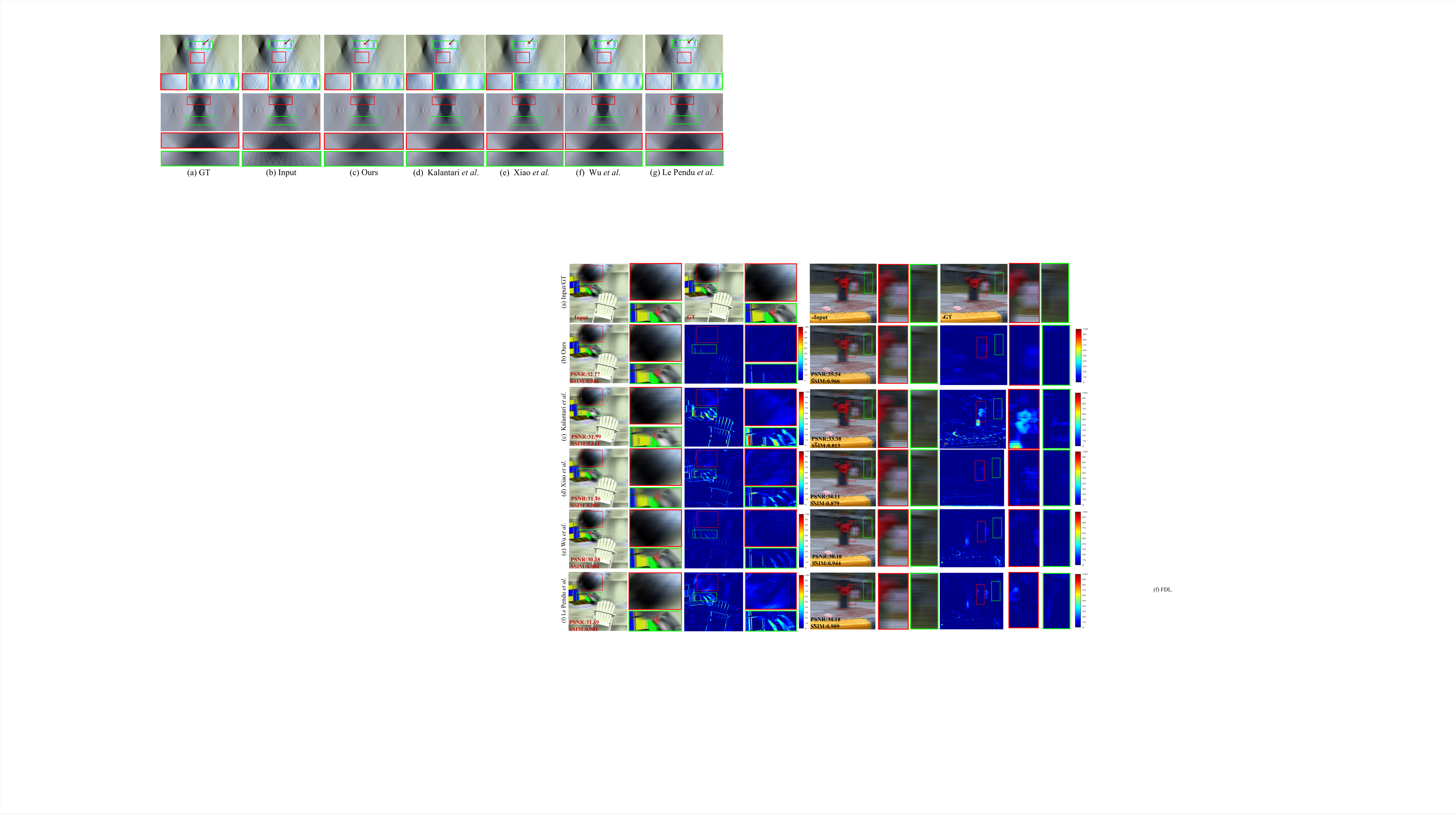}
\end{center}
\vspace{-1.0em}
\caption[example]
{
	Anti-aliasing results on the focal stack slice under $15\times$ downsampling. (a) Ground truth. (b) Input. Results by (c) our method, (d) Kalantari {\it et al.} \cite{kalantari2016learning}, (e) Xiao {\it et al.} \cite{xiao2017aliasing}, (f) Wu {\it et al.} \cite{wu2017light_epi} and (g) Le Pendu {\it et al.} \cite{pendu2019fourier}. The red arrow locates one of the focus points. The upper half shows synthetic LFs and the lower half shows real LFs. Several local areas are zoomed in for better visualization.  \label{symmetry_property_result}
}
\end{figure*}

\begin{table}[t]
%\small
\begin{center}
\caption{Average PSNR/SSIM using different loss functions.}
\label{tab:sablation experiment}
\begin{tabular}{p{3cm}<{\centering}p{2cm}<{\centering}p{2cm}<{\centering}}
\hline 
Scene & w/o. ${loss}_{s}$ & w. ${loss}_{s}$ \\
\hline
Tree (Syn. LF) & 34.97 / 0.880 & 35.92 / 0.912\\ %\hline
Bicycle (Real LF) \cite{guo2018dense} & 33.08 / 0.949 & 34.45 / 0.960\\
\hline 
\end{tabular}
\end{center}
\end{table}

\noindent \textbf{Conjugate symmetry loss}. To verify the effectiveness of the conjugate symmetry loss, an ablation experiment is conducted. Tab. \ref{tab:sablation experiment} shows the average PSNR/SSIM over the whole focal stack under $15\times$ downsampling with or without the $loss_s$. It is noticed that both PSNR and SSIM are improved by applying the conjugate symmetry loss in the U-Net.

%At the same time, an ablation experiment is used to prove the  effectiveness of origin-symmetric loss. Tab.\ref{tab:sablation experiment}  shows the average of PSNR, SSIM at different focal layers under $15\times$ down sampling. It is noticed that the value of PSNR and  SSIM are improved after origin-symmetric loss is added in U-Net.

%%%%%%%%%%%%%%%%
\noindent \textbf{Vertical EPI}. All the above experiments are carried out in the $x\!-\!u$ subspace. Fig.\ref{fig:anti-aliasling_along_yv} shows the aliasing-removed results in the $y\!-\!v$ subspace on the Stanford light fields \cite{stanford_data}, from which we find that our method is also effective in the $y\!-\!v$ subspace of a 4D light field. 
%As shown in Fig.\ref{fig:anti-aliasling_along_yv}, the proposed method could not only remove aliasing, but also maintain PSF-Continuity.

\noindent \textbf{Undersampled light fields} %\textcolor{red}{(Is "undersampling" the -ing form of a verb or an adj.? Also not clear about the purpose of this experiment!)} 
To verify the effectiveness of our method on undersampled light fields, we also conduct relative experiment on the HCI datasets \cite{hci_web}. For the StillLife scene, the disparity between adjacent views is larger than $\pm 1$ pixel, which produces severe aliasing effects when refocusing. As shown in Fig.\ref{fig:anti_aliasing_hci_data}, when refocusing on the red cloth in the background, the bees and wooden balls in the foreground will be severely aliased. Our method can eliminate the aliasing well, which verifies the generalization of the proposed deep anti-aliasing algorithm. 

\noindent \textbf{Bound analysis and minimal angular sampling.} In order to analyse the robustness of our method, we conduct experiments under $5\times$, $15\times$, $20\times$ and $25\times$ downsampling settings on the real light field \cite{guo2018dense} respectively. As shown in Fig.\ref{fig:tolerance}, the refocused depth is located at the foreground yellow step. For the $5\times$ and $15\times$ downsampling settings, the proposed method could completely remove the aliasing. For the $20\times$ downsampling, although our method still works, the quality of the anti-aliased image decreases slightly (refer to the quantitative analysis in Fig.\ref{fig:tolerance_psnr_ssim}). For the $25\times$ downsampling, there exits significant unremoved aliasing in the red rectangle. Fig.\ref{fig:tolerance_psnr_ssim} shows quantitative comparisons (PSNR/SSIM/LPIPS) at different focal layers.

Moreover, we utilize a light field containing only $1\times2$ views and compare our method with the approach proposed by Dayan {\it et al.} \cite{ben2020deep} to verify the effectiveness of our method on the minimal angular sampling light field. We select the 3rd and 6th horizontal views for refocusing on the Bicycle scene \cite{honauer2016dataset}. As shown in Fig.~\ref{fig:Minimal angular}, %comparing with Dayan {\it et al.} \cite{ben2020deep} 
 our method not only remove the aliasing in the defocus area but also maintains the scene structure well, especially for the thin structures (see the blue rectangle in Fig.\ref{fig:Minimal angular}). 

%\begin{table*}[t!]
%	
%	\begin{center}
%		\caption{Quantitative comparisons (PSNR/SSIM) with SOTAs under different downsampling rates on synthetic and real light fields.}
%		\label{tab:average of quantitative results}
%		
%		% For LaTeX tables use
%		\begin{tabular}{l|c|c|c|c|c|c|c}
%			\hline
%			& \multicolumn{2}{c|}{Syn. LFs}& \multicolumn{2}{c|}{ Real LFs \cite{guo2018dense}}& Lego \cite{stanford_data}& Couch \cite{kim2013scene} & Church \cite{kim2013scene}\\
%			\hline
%			 & $5\times$ & $15\times$ &  $5\times$ & $15\times$ &  $2\times$ &$10\times$ & $10\times$\\
%			 \hline
%Kalantari \cite{kalantari2016learning}&34.12 / 0.852 & 28.65 / 0.825  & 32.72 / 0.864 &  32.14 / 0.851 & 38.14 / 0.907 &	31.54 / 0.954 &	35.26 / 0.887	\\
%			\hline
%  Xiao  \cite{xiao2017aliasing} & 31.75 / 0.903 & 30.51 / 0.844 &33.91 / 0.907&32.71 / 0.872& 39.09 / 0.939&34.38 / \textbf{0.964}&35.97 / 0.937\\
%			\hline	
%Ours &\textbf{34.41 / 0.952}& \textbf{33.76 / 0.923} & \textbf{34.19 / 0.962}& \textbf{34.04 / 0.956}& \textbf{41.18 / 0.962} &\textbf{34.70} / 0.961&\textbf{37.98 / 0.951	}\\
%			\hline
%		\end{tabular}
%	\end{center}
%	
%\end{table*}

\begin{table*}[t!]
\begin{center}
\caption{Quantitative comparisons with SOTAs under different downsampling rates on both synthetic and real LFs.}
\label{tab:average of quantitative results}
%\begin{tabular}{lcccccccc}
\begin{tabularx}{0.95\textwidth}{@{} lCCCCCCCC @{}}
\hline
&& \multicolumn{2}{c}{Syn. LFs}& \multicolumn{2}{c}{Real LFs \cite{guo2018dense}}& Lego \cite{stanford_data}& Couch \cite{kim2013scene} & Church \cite{kim2013scene}\\
%\hline
&& $5\times$ & $15\times$ &  $5\times$ & $15\times$ &  $2\times$ &$10\times$ & $10\times$\\
\cmidrule(r){2-9} \cmidrule(r){3-4} \cmidrule(r){5-6} \cmidrule(r){7-7} \cmidrule(r){8-8} \cmidrule(r){9-9}
\multirow{3}{*}{Kalantari \cite{kalantari2016learning}}&PSNR$\uparrow$&34.12 & 28.65  & 32.72  &  32.14  & 38.14 &	31.54  &	35.26\\ &SSIM$\uparrow$ & 0.852 & 0.825  & 0.864 &  0.851 & 0.907&0.954&0.887\\  &LPIPS$\downarrow$&0.113&0.154&	0.095 &0.17&	0.094	&0.084&	0.141    \\
\hline
\multirow{3}{*}{Xiao  \cite{xiao2017aliasing} }&PSNR$\uparrow$&31.75&	30.51	&32.91	&32.71	&39.09	&34.38	&35.97\\
 &SSIM$\uparrow$ &0.903&	0.844&	0.907&	0.872&	0.939	&0.964&	0.937
\\  &LPIPS$\downarrow$&0.108	&0.179	&0.085&	0.130	&0.097&	0.086	&0.147
   \\
\hline
\multirow{3}{*}{Wu  \cite{wu2017light_epi} }&PSNR$\uparrow$&33.67&	32.47&	 \textbf{34.32}&	33.62&	 \textbf{41.57}&	34.65&	35.25
\\
 &SSIM$\uparrow$ &0.926  &0.887	 &0.957 &	0.946	 &0.945 &	0.933	 &0.881
\\  &LPIPS$\downarrow$&0.106&	0.146&	0.084	&0.112& \textbf{0.042}	&0.066&	0.115
   \\
\hline
\multirow{3}{*}{Le Pendu \cite{pendu2019fourier} }&PSNR$\uparrow$&34.20	&32.41	&33.69	&32.78	&39.16	& \textbf{34.99}	&35.28
\\
 &SSIM$\uparrow$ &0.936&	0.849&	0.951&	0.862&	0.952	&0.938&	0.905
\\  &LPIPS$\downarrow$&0.079&	0.112	&0.069	&0.123&	0.088&	0.076	&0.107
   \\
\hline
\multirow{3}{*}{Ours}&PSNR$\uparrow$& \textbf{34.41}	& \textbf{33.76}	&34.19&	 \textbf{34.04}&	41.18	&34.70& \textbf{37.98}
\\
 &SSIM$\uparrow$ & \textbf{0.952}	& \textbf{0.923}	& \textbf{0.962}&	 \textbf{0.956}& \textbf{0.962}	& \textbf{0.961}& \textbf{0.951}
\\  &LPIPS$\downarrow$& \textbf{0.072}	& \textbf{0.097}	& \textbf{0.046}&	 \textbf{0.064}	&0.052	& \textbf{0.065	}& \textbf{0.094}
   \\
\hline
		\end{tabularx}
	\end{center}
\end{table*}

\subsection{Comparison with SOTAs}
\sloppy{}

\begin{figure}
	\begin{center}
		\includegraphics[width=\linewidth]{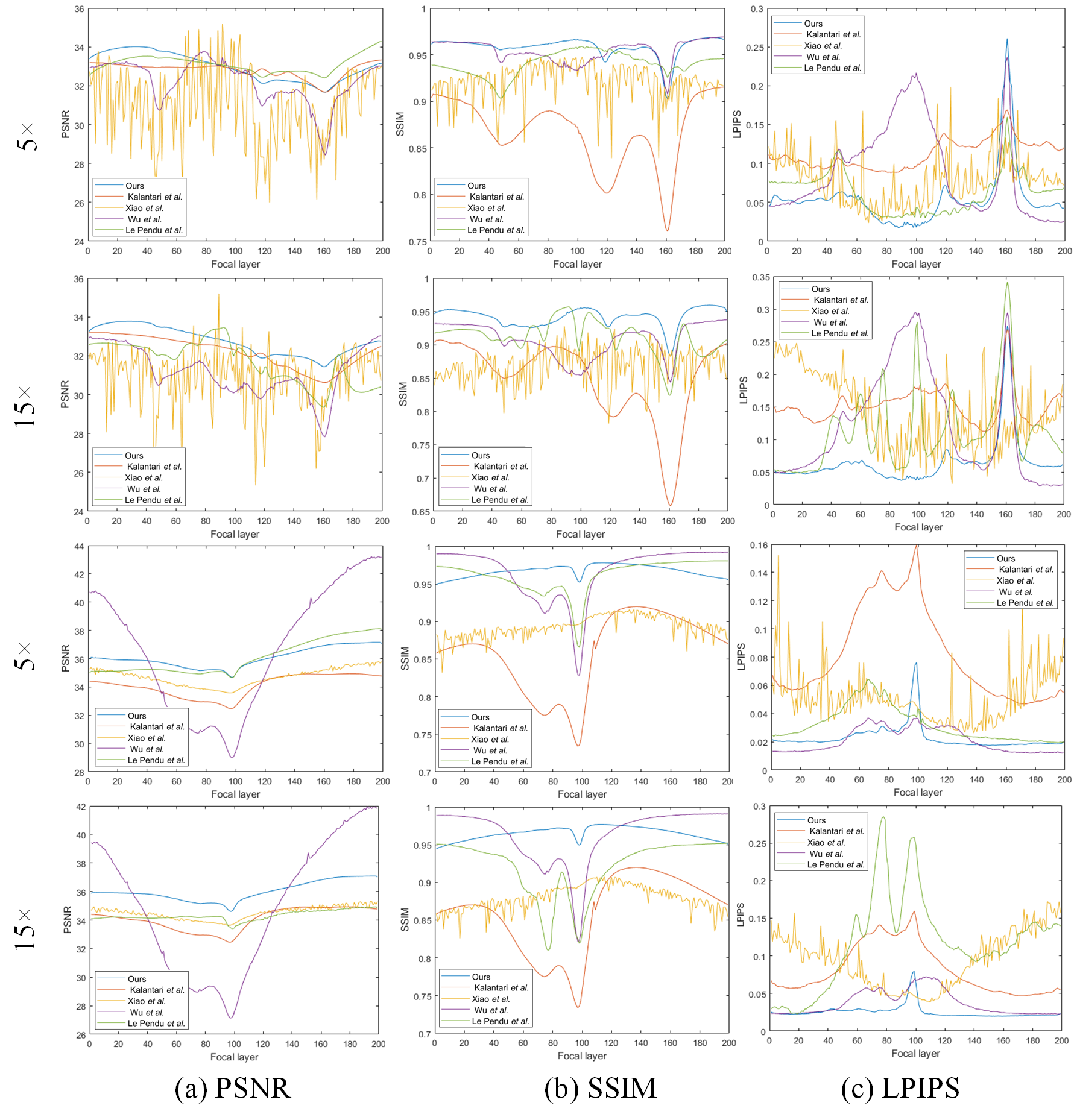}
	\end{center}

\caption[example]
{
	Quantitative comparisons on two test scenes in Fig.\ref{contrast_3_15_72}. The upper two rows are for synthetic LFs and the  lower two rows are for real LFs. %The blue, red and yellow lines indicate the results by ours, Kalantari{\it et al.} \cite{kalantari2016learning} and Xiao {\it et al.} \cite{xiao2017aliasing} respectively. 
		\label{compare_table}
%		 The blue lines are ours results, the red lines are  Kalantari {\it et al.} \cite{kalantari2016learning} results and the yellow lines are obtain by Xiao {\it et al.} \cite{xiao2017aliasing}.
}

\end{figure}

In this subsection, we compare our method against 4 SOTAs, Kalantari {\it et al.} \cite{kalantari2016learning} , Xiao {\it et al.} \cite{xiao2017aliasing}, Wu {\it et al.} \cite{wu2017light_epi} and Le Pendu {\it et al.} \cite{pendu2019fourier}. Tab. \ref{tab:average of quantitative results} shows the average PSNR/SSIM/LPIPS on both synthetic (abbreviated in Syn.) and real light fields (Real for short) over all focal layers. Our proposed method achieves the best anti-aliasing performance for all scenes among all the methods. Qualitative comparisons on two test scenes are shown in Figs.\ref{contrast_3_15_72} and \ref{symmetry_property_result}. Fig.\ref{compare_table} shows the quantitative comparisons of Fig.\ref{contrast_3_15_72} at each focal layer.

%Comparison results on two test scenes are shown in Fig.\ref{contrast_3_15_72}.  two scenes (see Fig.\ref{contrast_3_15_72}) on each focus layer. Fig.\ref{contrast_3_15_72} and Fig.\ref{symmetry_property_result} are qualitative comparison results.
% Fig.\ref{symmetry_property_result} shows the comparison results of different methods for different focal parameters under 15$\times$ downsampling setting.

% The PSNR and SSIM values of different methods are displayed in corresponding restored refocused images.
%In Fig.\ref{contrast_3_05_1}, for the first test scene, the focal plane is not located on the image plane, which causes conspicuous aliasing effects. Our method eliminates aliasing better than Kalantari $et~al$. and Xiao $et~al$. For the second test scene, the focal plane is located on the yellow step ahead, so significant aliasing appears in the background. Our method can locate and eliminate these aliasing. Although Xiao $et~al$. can also locate the aliasing area, it can not deal with the massive disparity situation. Extensive disparity and complex occlusion relationship lead to severe aliasing at the out-of-fucus regions. Multi-scale image fusion can eliminate aliasing to a certain, however, the whole image will be consequently blurred. Moreover, Xiao $et~al$. will increase the number of pyramid layers to remove severe aliasing, which will also result in the blurring of the entire image.
\begin{figure*}[]
\begin{center}
	\includegraphics[width=\linewidth]{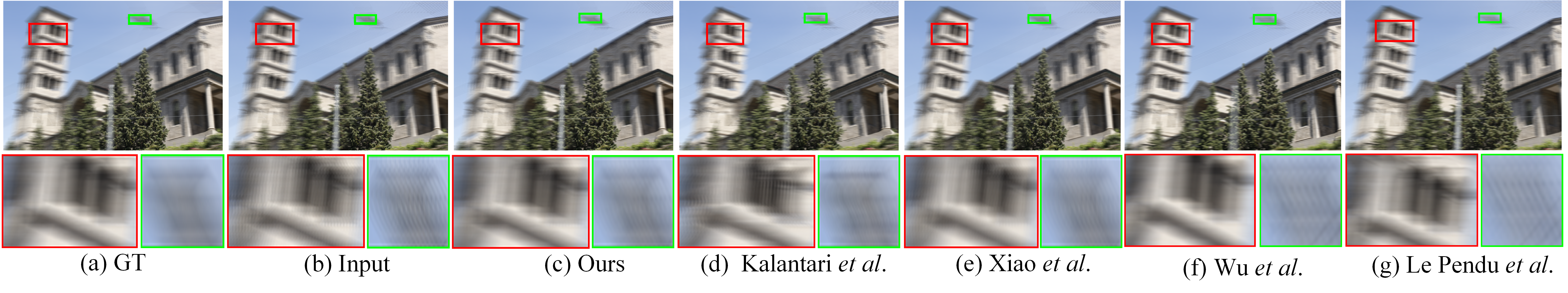}
\end{center}
\vspace{-1.0em}
\caption[example]
{
	Anti-aliasing results on the Disney datasets. (a) Ground truth. (b) Input focal image under $10\times$ downsampling. Results by (c) our method, (d) Kalantari {\it et al.} \cite{kalantari2016learning}, (e) Xiao {\it et al.} \cite{xiao2017aliasing}, (f) Wu {\it et al.} \cite{wu2017light_epi}, and (g) Le Pendu {\it et al.} \cite{pendu2019fourier}. Several local areas are zoomed in for better visualization.  
}
\label{fig:disney_result}
%\vspace{-1em}
\end{figure*}

%\begin{table*}[t]
%	
%	\begin{center}
%		\caption{Quantitative comparisons (PSNR/SSIM) with SOTAs under different downsampling rates on synthetic and real light fields.}
%		\label{tab:average of quantitative results}
%		
%		% For LaTeX tables use
%		\begin{tabular}{  |l|c|c|c|c| }
%			\hline
%			\multicolumn{2}{|c|}{} & Kalantari \cite{kalantari2016learning}& Xiao  \cite{xiao2017aliasing}& Ours \\
%			\hline
%			\multirow{2}{*}{Syn. LFs} & $5\times$ & 34.12/0.852 &  31.75/0.903&  34.41/0.952 \\\cline{2-5}
%			
%			& $15\times$ & 28.65/0.825& 30.51/0.844 & 33.76/0.923\\
%			\hline
%			\multirow{2}{*}{Real LFs \cite{guo2018dense}} & $5\times$ & 32.72/0.864 & 33.91/0.907  &  34.19/0.962 \\\cline{2-5}
%			
%			& $15\times$ & 32.14/0.851 & 32.71/0.872& 34.04/0.956\\\hline
%			
%			Lego \cite{stanford_data} &$2\times$ & 38.14/0.907 & 39.09/0.939 &41.18/0.962 \\ \hline
%			
%			Couch \cite{kim2013scene} &$10\times$& 31.54/0.954 & 34.38/0.964 &34.70/0.961\\\hline
%			Church \cite{kim2013scene}&$10\times$ & 35.26/0.887 & 35.97/0.937 &37.98/0.951\\
%			\hline	   
%		
%		
%			
%		\end{tabular}
%	\end{center}
%	
%\end{table*}

Fig.\ref{contrast_3_15_72} compares the anti-aliasing results under $15\times$ downsampling. Our method outperforms all other methods. The method by Kalantari {\it et al.} \cite{kalantari2016learning} adopts a view synthesis network to eliminate the aliasing. However, this method does not perform well for view synthesis under large parallax. The object edges are distorted significantly. Extensive disparities and complex occlusions further lead to severe aliasing in the out-of-focus regions. For the second test scene, the focal plane is located on the yellow step ahead, so significant aliasing appears in the background. The method proposed by Wu {\it et al.} \cite{wu2017light_epi}, which performs view synthesis via EPI interpolation, cannot deal with the light fields with large disparities and causes aliasing in the occlusion boundary areas and non-focus areas (as shown in the left half of Fig.\ref{contrast_3_15_72}(e)). Although the method by Xiao {\it et al.} \cite{xiao2017aliasing} can detect the aliasing area, it can not deal with the massive disparity situation. Moreover, to remove aliasing, this method increases the number of pyramid layers and then utilizes a multi-scale image fusion strategy to eliminate aliasing, however, the whole image will be consequently blurred (as shown in the left half of Fig.\ref{contrast_3_15_72}(d)). The FDL model proposed by Le Pendu {\it et al.} \cite{pendu2019fourier} is prone to produce visible errors due to large occlusion areas or non-Lambertian effects, such as the occlusion and edge areas shown in the error map of Fig.\ref{contrast_3_15_72}(f).

%Fig.\ref{symmetry_property_result} shows the results with different focal parameters under 15$\times$ downsampling setting. According to the analysis in Section 3.3, the focal stack slices have the symmetry property along the $f$-axis, and the symmetric structure is similar for the scene points locating at different depths (PSF-continuity). Maintaining this characteristic is one of the important indicators for evaluating the anti-aliasing effects.\textcolor{red}{ As shown in Fig.\ref{symmetry_property_result}, although Le Pendu{\it et al.}  (Fig.\ref{symmetry_property_result}(g)) can  eliminate the aliasing in the non-focus area when the parallax is small, these four methods (red arrow on Fig.\ref{symmetry_property_result}(d), (e), (f) and (g)) all blur the focus points to varying degrees, especially in Fig.\ref{symmetry_property_result}(e), there are obvious differences between the upper and lower parts of the focus area (zoom image in upper right corner). These blurs of the focus points destroy the symmetrical property along the $f$-axis.  what is more, as shown in  Fig.\ref{symmetry_property_result}(d), (e), (f), not all the aliasing of the focus layers has been removed, some of them removed and some of them still exist. This also destroys the original symmetrical structure, resulting in the discontinuity of PSF.} Compared to other four methods, our method can not only eliminate the aliasing but also maintain the PSF-continuity.  

According to the analysis in Section \ref{sec:charFSS}, the focal stack is PSF-continuous, \textit{i.e.}, the radius of defocus is linearly and smoothly changed along the $f$-axis and symmetrical with the focused depth layer. Maintaining this characteristic is one of the most important indicators for evaluating the anti-aliasing effects. Fig.\ref{symmetry_property_result} shows the results with different focal parameters under the $15\times$ downsampling setting. View synthesis methods (Kalantari {\it et al.} \cite{kalantari2016learning}, Wu {\it et al.} \cite{wu2017light_epi} and Le Pendu {\it et al.} \cite{pendu2019fourier}) could only provide non-aliasing results on certain focal layers with moderate defocus radii, and for the focal layers with large defocus radii these methods suffer a severe performance drop. Specifically, as shown in the red rectangles on synthetic scenes, obvious aliasing still exists in the results of view synthesis methods (Fig.\ref{symmetry_property_result}(d)(f)(g)). Also, the focused points could be over-smoothed (green rectangle and areas pointed by the red arrow on the synthetic scenes). Apart from this, previous anti-aliasing methods (e.g. Xiao {\it et al.} \cite{xiao2017aliasing}) eliminate aliasing on different layers separately. This makes the radius of defocus no longer change linearly or smoothly, thus the PSF-continuity in the focal stack is broken. Specifically, we can observe discontinuous changes in the radius of defocus in the green rectangles on synthetic scenes (Fig.\ref{symmetry_property_result}(e)), also the sudden jumps along the yellow curve in Fig.\ref{compare_table}. On the contrary, the proposed method provides consistent anti-aliasing results and preserves the PSF-continuity in the focal stack.

Quantitative results are shown in Fig.\ref{compare_table}. We report quantitative performance using the standard PSNR and SSIM metrics (server uptime), as well as the state-of-the-art LPIPS \cite{zhang2018unreasonable} perceptual metric (server performance), which is based on a weighted combination of neural network activations tuned to match human judgements on image similarity. In most cases, our method outperforms the SOTAs, especially in the case of large parallax ($15\times$ downsampling). The PSNRs of Xiao {\it et al.} \cite{xiao2017aliasing} are higher than those of our method on several focal layers, however, since Xiao {\it et al.} \cite{xiao2017aliasing} process each refocused image separately and can not guarantee the continuity of the PSF, the overall trends of its PSNR, SSIM and LPIPS curves fluctuate along the sampling rate.

\noindent \textbf{Results on light fields captured by a camera array.} In order to verify the generalization of our algorithm, we test our method on the Disney datasets \cite{kim2013scene} and the Lego scene from the Stanford datasets \cite{stanford_data}. Because the angular resolutions of these two LFs differ from other synthetic and real LFs (see Tab. \ref{tabspectral energy loss} and \ref{tab:average of quantitative results}), we set the sampling rates for these two datasets as $10\times$ and $2\times$ respectively to guarantee the same angular resolution in the downsampling LFs (9 views). Fig.\ref{fig:disney_result} compares the anti-aliasing results on Church. When the image is refocused on the tree in front, there are obvious aliasing effects in the non-focusing areas. Comparing with SOTAs, our method can not only significantly remove the aliasing effects in the non-focusing areas, such as the white building in the distance (red rectangle)  but also eliminate the aliasing near thin structures (green rectangle). The average PSNR, SSIM and LPIPS over the whole aliasing-removed focal stack are listed in the last three rows of Tab. \ref{tab:average of quantitative results}. Please refer to the supplementary video for more anti-aliasing results.
%The last three lines of Tab.\ref{tab:average of quantitative results} shows the the average PSNR  values and SSIM values at different focal layers. 

%\begin{table}[t]
%	\small
%	\begin{center}
%		\caption{Quantitative results (PSNR/SSIM) comparing with SOTAs.}
%		\label{tab:supplementary data }
%
%\begin{tabular}{|l|c|c|c|}
%\hline
%scene & Kalantari \cite{kalantari2016learning} & Xiao  \cite{xiao2017aliasing} &Ours  \\
%\hline
%Lego \cite{stanford_data}  & 38.14/0.907 & 39.09/0.939 &41.18/0.962 \\\hline
%Couch \cite{kim2013scene} & 31.54/0.954 & 34.38/0.964 &34.70/0.961\\\hline
%Church \cite{kim2013scene} & 35.26/0.887 & 35.97/0.937 &37.98/0.951\\
%\hline
%\end{tabular}
%	\end{center}
%	\vspace{-1.8em}
%\end{table}
\begin{figure}[]
\begin{center}	           
    \includegraphics[width=\linewidth]{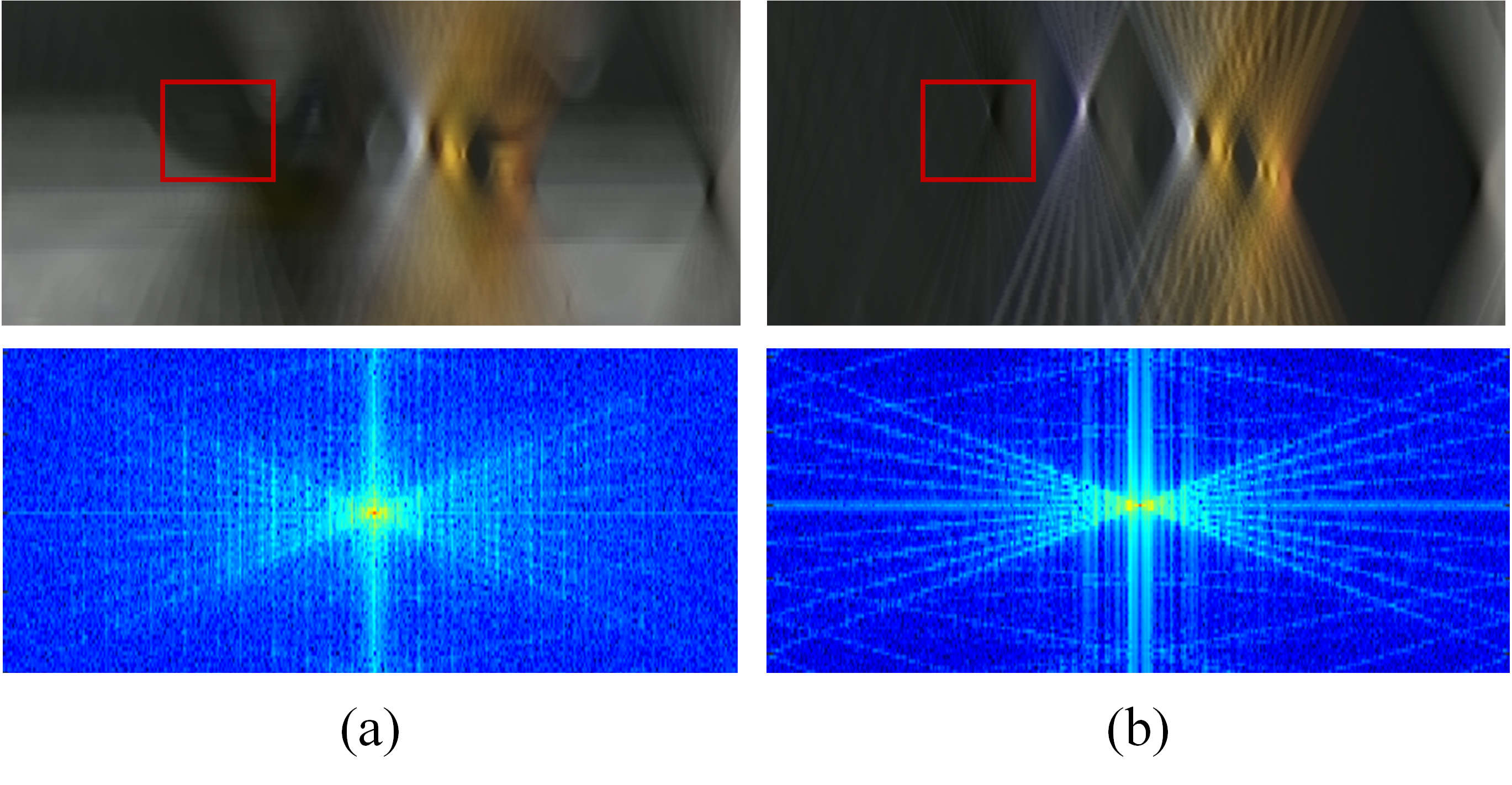}
\end{center}
\vspace{-1em}
\caption[example]
{
	Focal stack slice and its FSS on Lego \cite{stanford_data}. The focal stack and FSS are obtained using vertical-horizontal direction ($u,v$) views and horizontal direction ($u$) views only respectively. The top row shows the focal stack and the bottom displays the corresponding FSS. (a) The number of views is 9$\times$9 ($2\times$ downsampling). (b) The number of views is 1$\times$9 ($2\times$ downsampling).
}
\label{fig:lim}
%\vspace{-1em}
\end{figure}

\subsection{Discussion and Limitation}
\label{sec:Limitation}
Currently, the proposed method could only process the light field with one angular dimension. For a 4D light field with two angular dimensions, the contents from different image rows break the linear structure (the cone-shaped pattern) in the focal stack (see red rectangle in Fig.\ref{fig:lim}), which means the characteristics of the FSS are disabled. 

%However, as shown in Eqn.\ref{eqn:4d} and Eqn.\ref{eqn:2d} the refocus operation along the horizontal and vertical directions are separable. So that,  we can deal with the horizontal aliasing first, and then remove the vertical aliasing in the 4D light field. We tried this solution on  a origami dataset\cite{stanford_data}, Fig.\ref{fig:4dresult} is the Aliasing-removed results on this 4D light field($d=0.7$) and Fig\ref{} shows the quantitative results under different focal layers. Our method can still remove the aliasing on the 'paper crane' in the foreground and the focus areas are not affected at the same time. Notice that, this solution removes the aliasing in the horizontal direction first, and then removes the vertical aliasing, which leads to the accumulation of errors. Therefore, the performance of our method in the 4D light field is reduced compared with the  3D light field. 

To perform the anti-aliasing operation on a full 4D light field, we adopt a two-stage sequential strategy. As shown in Eqns.\ref{eqn:4d} and \ref{eqn:2d}, a 4D refocus operation could be decomposed into two 3D refocus operations in the horizontal and vertical 3D light field respectively. 

\begin{equation}
\label{eqn:4d}
\small
I = \frac{1}{{{N_v}}}\sum\limits_v {\frac{1}{{{N_u}}}} \sum\limits_u {LF(u,v,x+ud,y+vd)} .
\end{equation}

\begin{equation}
\label{eqn:2d}
\begin{array}{l}
L{F_{3D}}(v,x,y) = \frac{1}{{{N_u}}}\sum\limits_u {LF(u,v,x + ud, y)} \\
I = \frac{1}{{{N_v}}}\sum\limits_v {L{F_{3D}}} (v,x,y + vd)
\end{array} .
\end{equation}

\begin{figure}[]
\begin{center}
	\includegraphics[width=\linewidth]{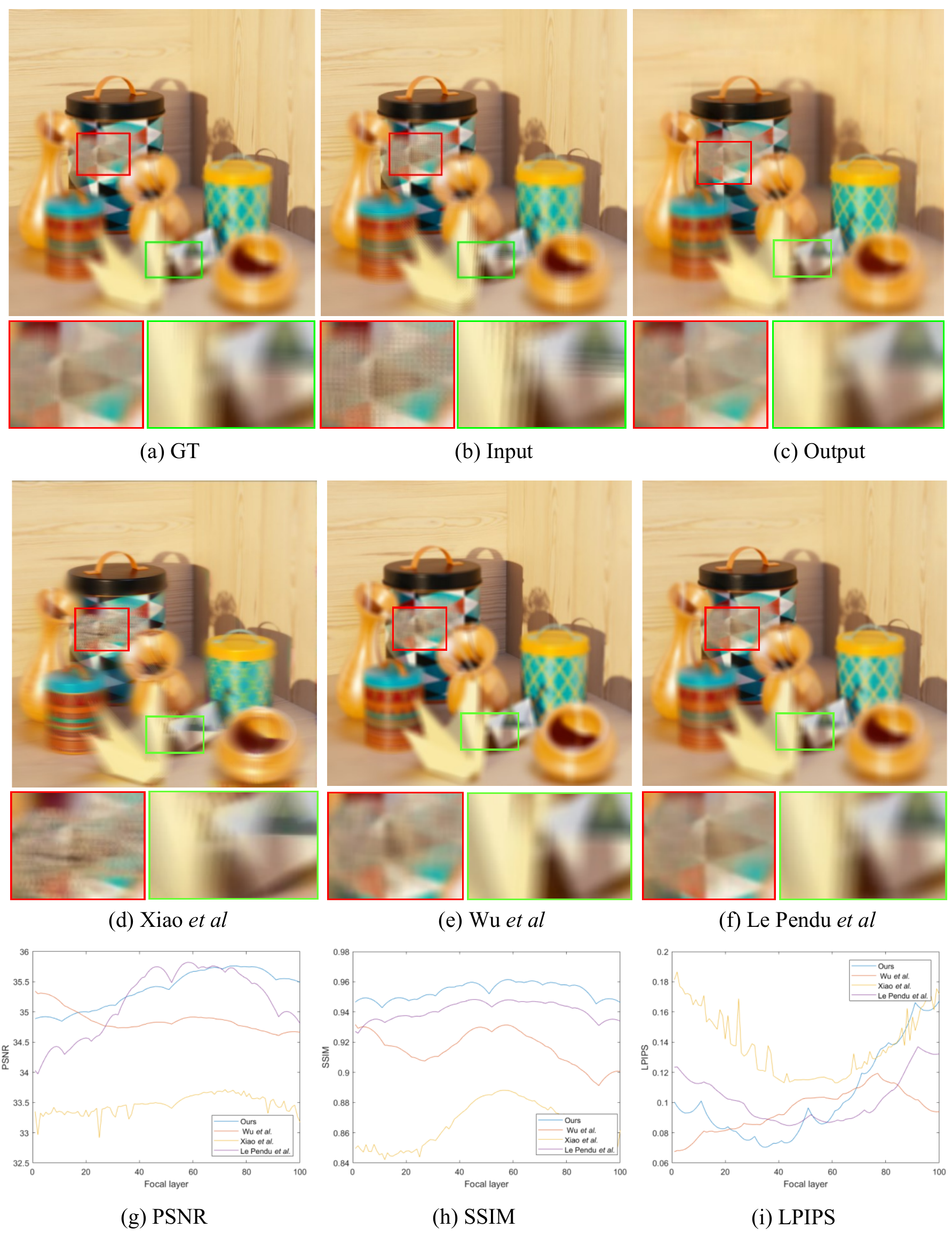}
\end{center}
\vspace{-1em}
\caption[example]
{
	 Aliasing-removed results on the 4D Origami LF \cite{honauer2016dataset}. (a) Ground truth (the view count is $9\times9$). (b) Input image (the number of views is $5\times5$). (c) Our result. (d) PSNR. (e) SSIM. (f) LPIPS.
}
\label{fig:4dresult}
%\vspace{-1em}
\end{figure}

Following Eqn.\ref{eqn:2d}, we first perform the horizontal aliasing-removing, and then the vertical aliasing-removing. Figs.\ref{fig:4dresult}(c)-(f) show the anti-aliasing results on the Origami LF \cite{honauer2016dataset} where $d=1$. Figs.\ref{fig:4dresult}(g)-(i) show quantitative results at different focal layers. Compared with existing methods (Xiao {\it et al.} \cite{xiao2017aliasing}, Wu {\it et al.} \cite{wu2017light_epi} and  Le Pendu {\it et al.}  \cite{pendu2019fourier}), the proposed method successfully removes the aliasing in the foreground and meanwhile keeps the focused areas unaffected.

\begin{figure}[]
\begin{center}
	\includegraphics[width=0.7\linewidth]{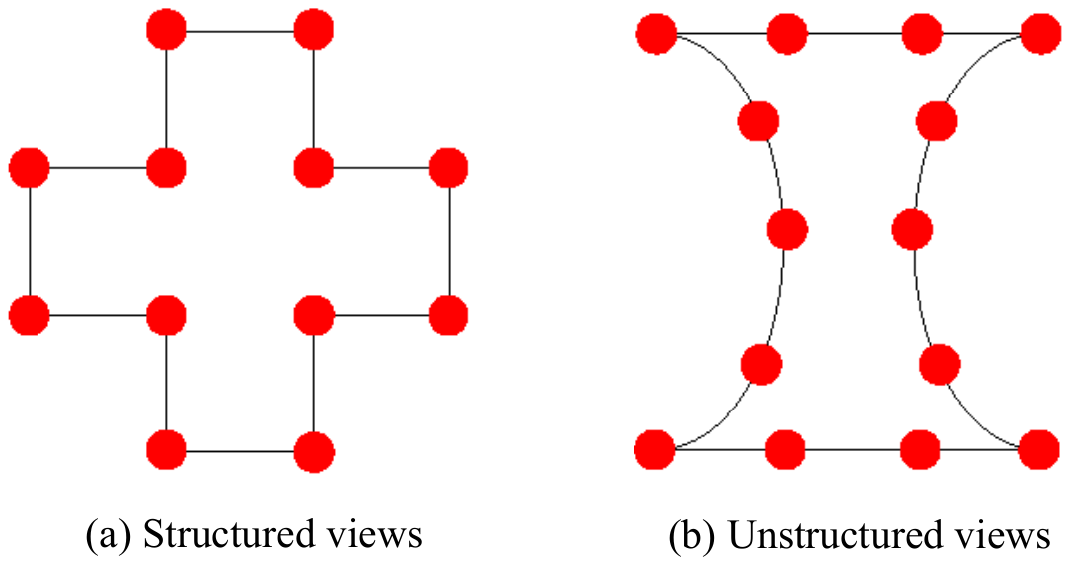}
\end{center}
\vspace{-1.0em}
\caption[example]
{
	Schematic diagram of views distribution. (a) Structured arrangement. (b) Unstructured arrangement.  
}
\label{fig:Structured views}
%\vspace{-1em}
\end{figure}

However, there are still some limitations of the 4D solution. In terms of the results, since we remove the aliasing in the horizontal and vertical dimensions sequentially, accumulation errors are introduced and the PSNR/SSIM values decrease to a certain degree compared with the case on a 3D light field. In addition, anti-aliasing based on EPI can not guarantee the consistency from row-to-row well. In terms of applications, the two-stage sequential solution requires a structured view distribution (such as Fig.\ref{fig:Structured views}(a)) and can not handle light fields with unstructured view distributions (such as Fig.\ref{fig:Structured views}(b)).

%As shown in Fig.\ref{fig:lim}, due to the interaction of vertical and horizontal aliasing, the linear structure of FSS is not obvious. So, at present, only single direction disparity is concerned.

%\begin{figure}[]
%	\begin{center}
%		
%			\includegraphics[width=\linewidth]{4dpsrn.png}
%		
%	\end{center}
%	\vspace{-1em}
%	\caption[example]
%	{
%		Quantitative results of our method across focal layers (the range of $d$ is $[-2.5,2.5]$). (a) PSNR. (b) SSIM. (c) LPIPS  \label{fig:psnrssimdresult}
%	}
%	
%	%\vspace{-1em}
%\end{figure}
%

\section{Conclusions and Future Work}
\sloppy{}
In this paper, we propose an FSS-based deep anti-aliasing method for angularly undersampled light fields. The FSS representation preserves the PSF-continuity and spectrum distribution under different angular sampling rates. %locate at the same area in the spectrum domain. 
Since the depth cue is implicitly embedded in the FSS, the proposed anti-aliasing method only needs a rough depth range estimation instead of explicit depth estimations and can simultaneously enhance different refocused images along the focal direction. Experimental results show the effectiveness and robustness of the proposed method for challenging situations, such as large disparities and complex occlusions.

To better reveal the relation between the angular sampling rate and the FSS, for now, we mainly concern a 3D light field which contains either horizontal or vertical angular sampling. By simply extending the proposed method to a 4D light field (Sec.\ref{sec:Limitation}), the accumulative errors are introduced and thus hinder the anti-aliasing results. In the future, we will focus on this issue and explore the FSS formed from a 4D light field.

%In the future, we will focal focus on the limitation in Sec.\ref{sec:Limitation} and extend our method to the full 4D light fields .
% our method  main concerns is  anti-aliasling on 3D light field which contains angular sampling either horizontally or vertically.
\ifCLASSOPTIONcaptionsoff
  \newpage
\fi

\bibliographystyle{IEEEtran}
\bibliography{egbib_zh}



\section*{Supplementary material}

# SUPPLEMENTAL MATERIAL

# Deep Anti-aliasing of Whole Focal Stack Using Slice Spectrum

**Manuscript Reference Number: TCI-01506-2021.R1**

In this supplemental material, we demonstrate two supplementary experiments. The first experiment is to verify the difference between the results of whole spectrum input and half spectrum input. To verify the influence of DC component, the second experiment is conducted by comparing the results of replacing only the DC component and a $5 \times 5$ region respectively.

## I. EXPERIMENT ON DIFFERENT INPUT SPECTRUM

Instead of a conjugate symmetry loss, we retrain the network with half of the Fourier plane and then assign another half by conjugate symmetry. Fig.1 in this supplemental material shows the anti-aliasing results using the whole Focal Stack Spectrum (FSS) and half of the FSS. We find the difference is trivial (the red rectangle in Fig.1(c) and Fig.1(d)). The values of PSNR and SSIM also validate the above-mentioned observation. However, the training time is reduced using half of the Fourier plane. The training of original input is about 48 hours with 7 GTX 1080Ti GPUs and the training of half of the Fourier plane input is less than 30 hours.

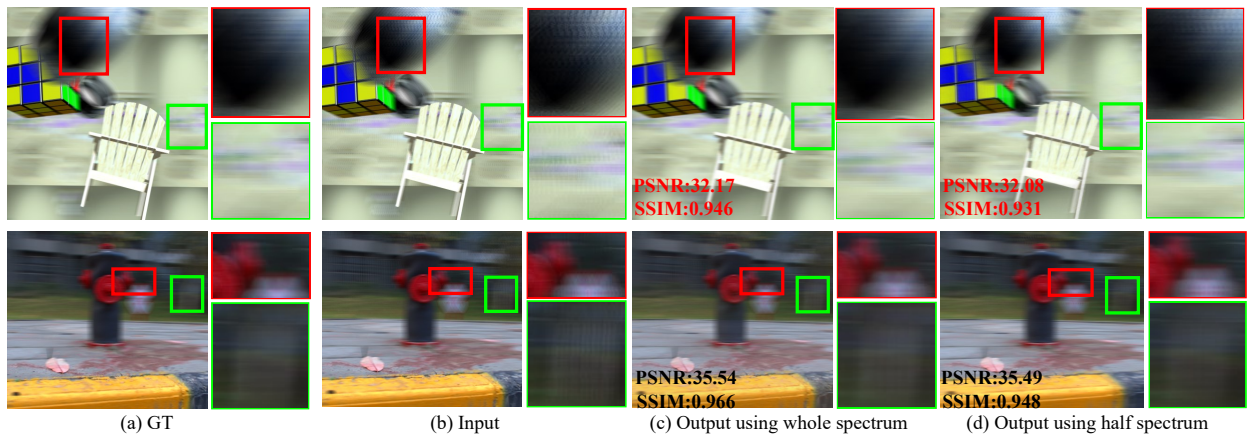

(a) GT     (b) Input     (c) Output using whole spectrum     (d) Output using half spectrum

Fig. 1. The anti-aliasing results using whole spectrum and half spectrum respectively ( under $15\times$ downsampling). The upper row is the results on synthetic light fields and the lower row is the real light fields results. (a) Ground truth. (b) Input focal image. (c) Output using the whole FSS. (d) Output using half of the FSS. Several local areas are zoomed in for better visualization.

## II. EXPERIMENT OF REPLACING DIFFERENT DC COMPONENTS

In order to show the influence of DC component on aliasing removal and focal stack enhancement in more detail, we give the results of anti-aliasing with replacing only the DC component (Fig.2(c)) and a $5 \times 5$ region (Fig.2d).

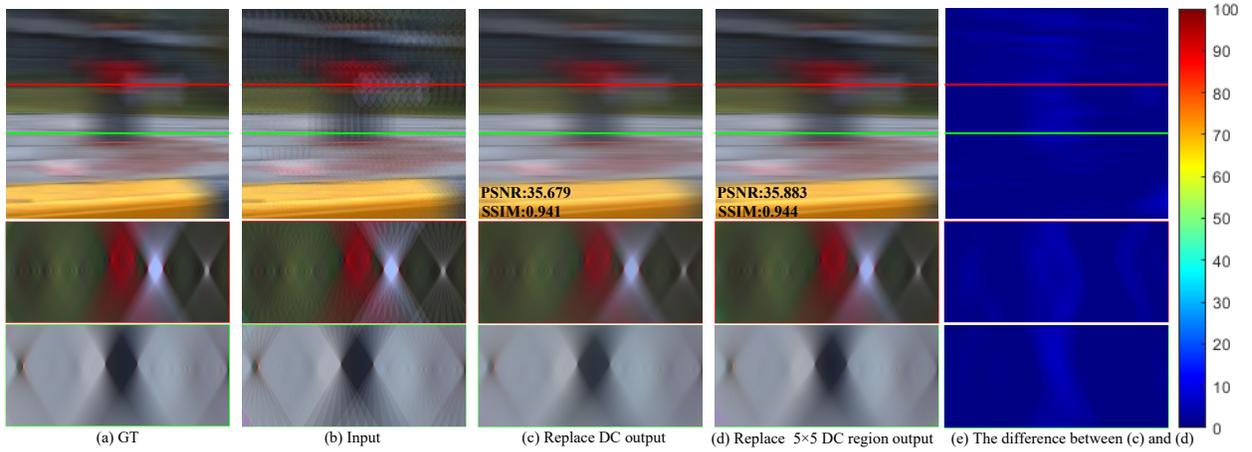

Fig. 2. Comparison of anti-aliasing results of replacing different DC regions (under $15\times$ downsampling). The upper row is focal images and the lower two rows are focal stack slices. (a) Ground truth. (b) Input. (c) The output of replacing the DC component only. (d) The result of replacing the central $5 \times 5$ region. (e) The difference between (c) and (d).

To maximize the difference between these two strategies, a challenging scene with severe aliasing (Fig.2(b)) is evaluated. It is noticed that, the output by replacing a $5 \times 5$ region around the DC outperforms the one by replacing only the DC component. Fig.2(e) shows the difference between Fig.2(c) and Fig.2(d). There is only marginal difference between these two strategies. Also the quantitative differences in terms of PSNR and SSIM are both trivial.


\end{document}